\documentclass[structabstract]{aa}
\usepackage{enumerate}

\usepackage{subfigure}
\usepackage{amsmath} 
\usepackage[version=3]{mhchem}
\usepackage{graphicx}
\usepackage{xcolor}
\usepackage{txfonts}
\usepackage{natbib}
\usepackage{float}
\begin{document}

\title{Photodissociation and chemistry of N$_2$ in the circumstellar envelope of carbon-rich AGB stars}


\author{Xiaohu Li\inst{1}
        \and T. J.\ Millar\inst{2}
        \and Catherine Walsh\inst{1}
        \and Alan N.\ Heays\inst{1}
        \and Ewine F. van Dishoeck\inst{1,3}
        }

        \institute{$^1$ Leiden Observatory, Leiden University,
          P.O. Box 9513, 2300 RA Leiden, The Netherlands\\
          $^2$ Astrophysics Research Centre, School of Mathematics and Physics, Queen's University Belfast, Belfast, BT7 1NN, UK \\
          $^3$ Max-Planck Institut f\"ur Extraterrestrische Physik
          (MPE), Giessenbachstr.\ 1, 85748 Garching, Germany \\
          \email{li@strw.leidenuniv.nl}}

\date{}

\titlerunning{Photodissociation and chemistry of N$_2$ in the circumstellar envelopes of carbon-rich AGB stars}
\authorrunning{Li et al.}




\def\placefigurecsestructure
{  
 
\begin{figure*}

\centering
\includegraphics[width=0.8\textwidth]{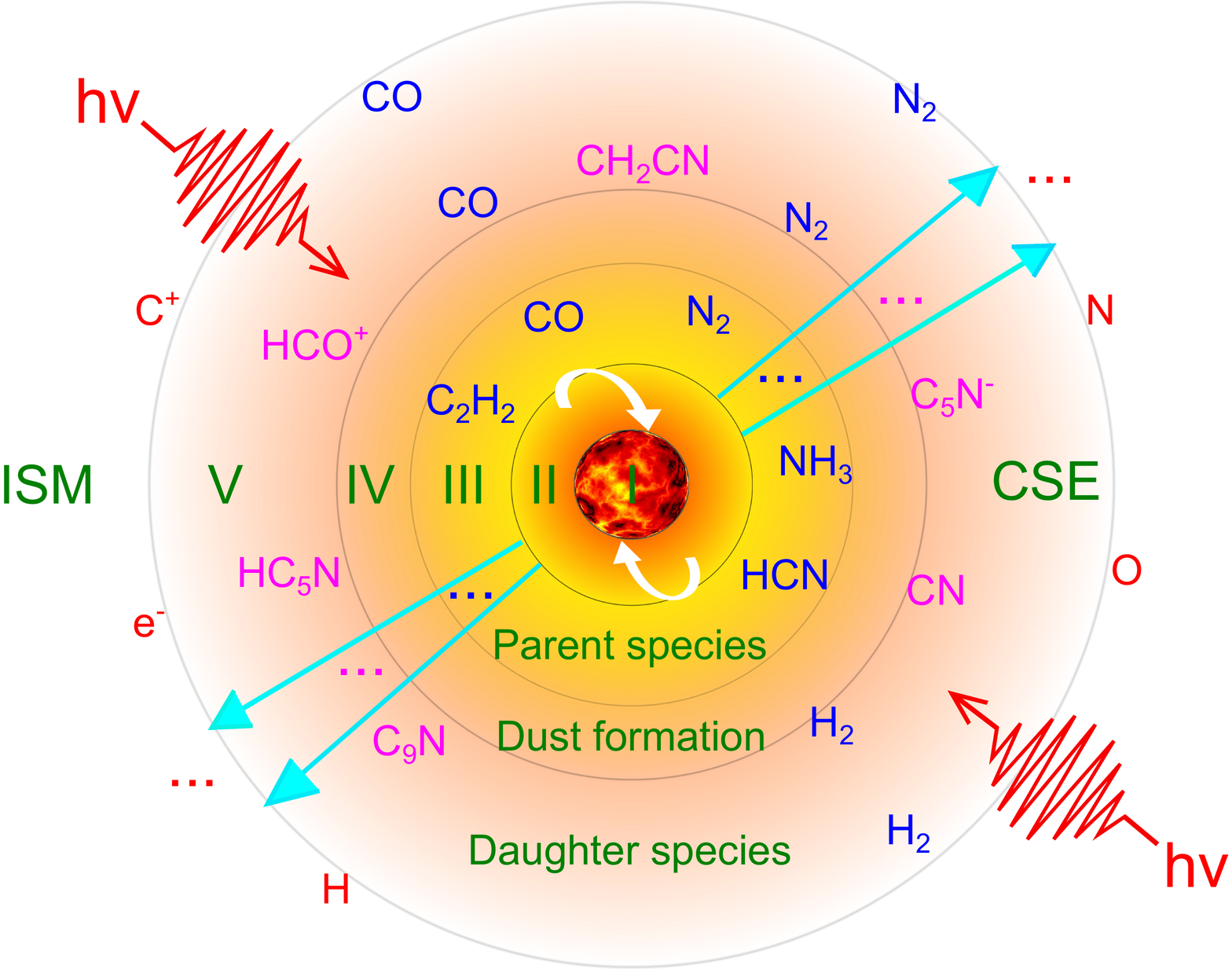}
\caption{Schematic structure of the CSE for a C-rich AGB star, which is divided into 6 regions for modelling purposes.
(\uppercase\expandafter{\romannumeral1}): a degenerate C/O core and He/H burning shell,
(\uppercase\expandafter{\romannumeral2}): a convective shell, 
(\uppercase\expandafter{\romannumeral3}): a stellar atmosphere in which parent species are formed,
(\uppercase\expandafter{\romannumeral4}): a dust formation shell with an expanding envelope,
(\uppercase\expandafter{\romannumeral5}): an outer CSE where daughter species are formed primarily by photodissociation, 
(\uppercase\expandafter{\romannumeral6}): the interstellar medium (ISM). 
This study focusses on the outer CSE where chemistry is mainly driven by the photodissociation of molecules. 
}

\label{figcsesturcture}  
\end{figure*}

}  

\def\placefigureLoop
{  
 
\begin{figure}

\centering
\includegraphics[width=0.45\textwidth]{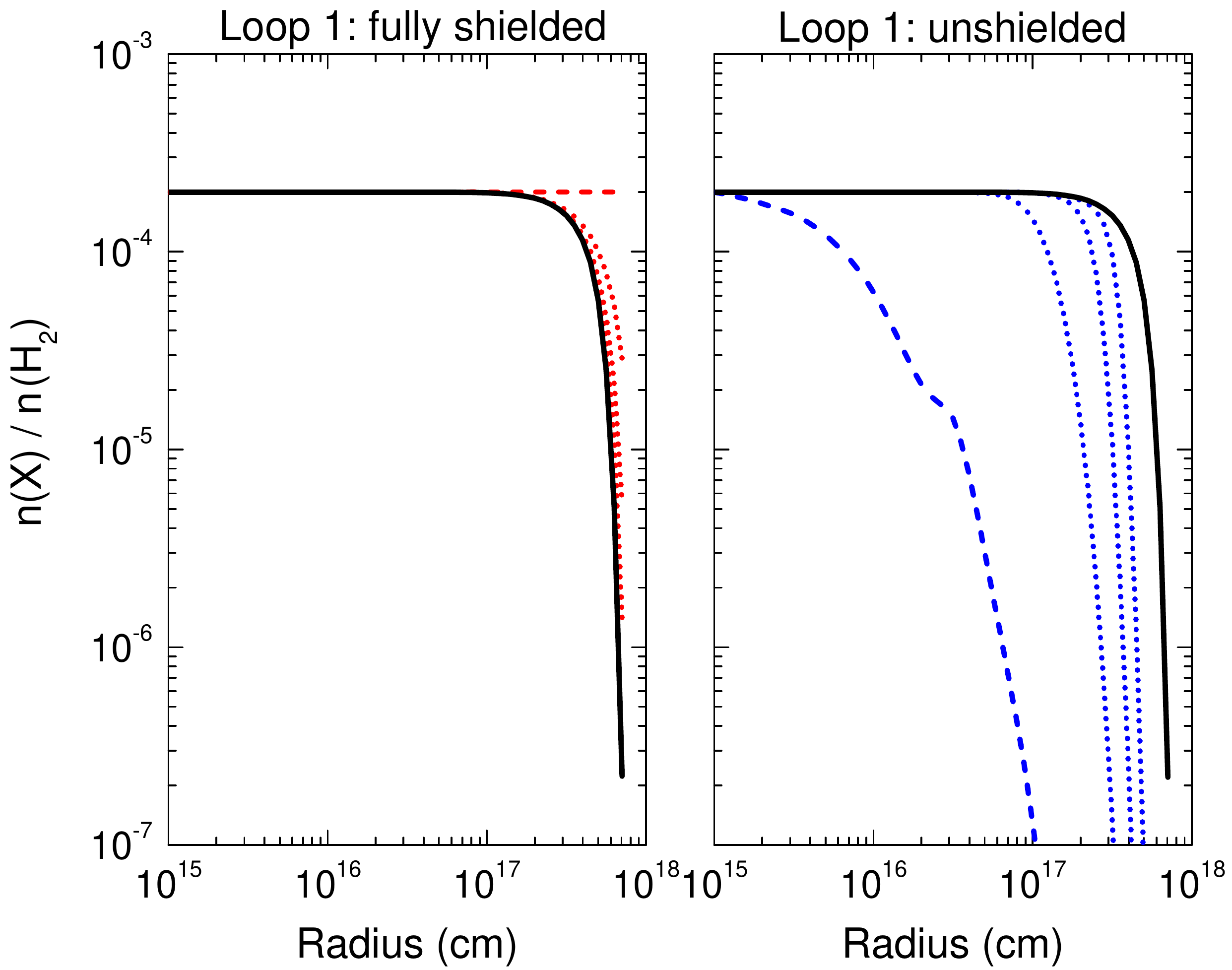}
\caption{Illustrations of the numerical method employed for implementing N$_2$ (and CO) shielding functions in this work. The plots show the fractional abundances of N$_2$ as functions of radii. It is seen that one can obtain the same converged results (solid lines in black) by assuming N$_2$ is either fully shielded (dashed line in red, left panel) or unshielded (dashed line in blue, right panel) at the first loop. The dotted lines show the intermediate abundances before reaching the final results.  
}
\label{figloop}  
\end{figure}

}   

\def\placefigureT
{  
 
\begin{figure}
\centering
\includegraphics[angle=0,width=0.45\textwidth]{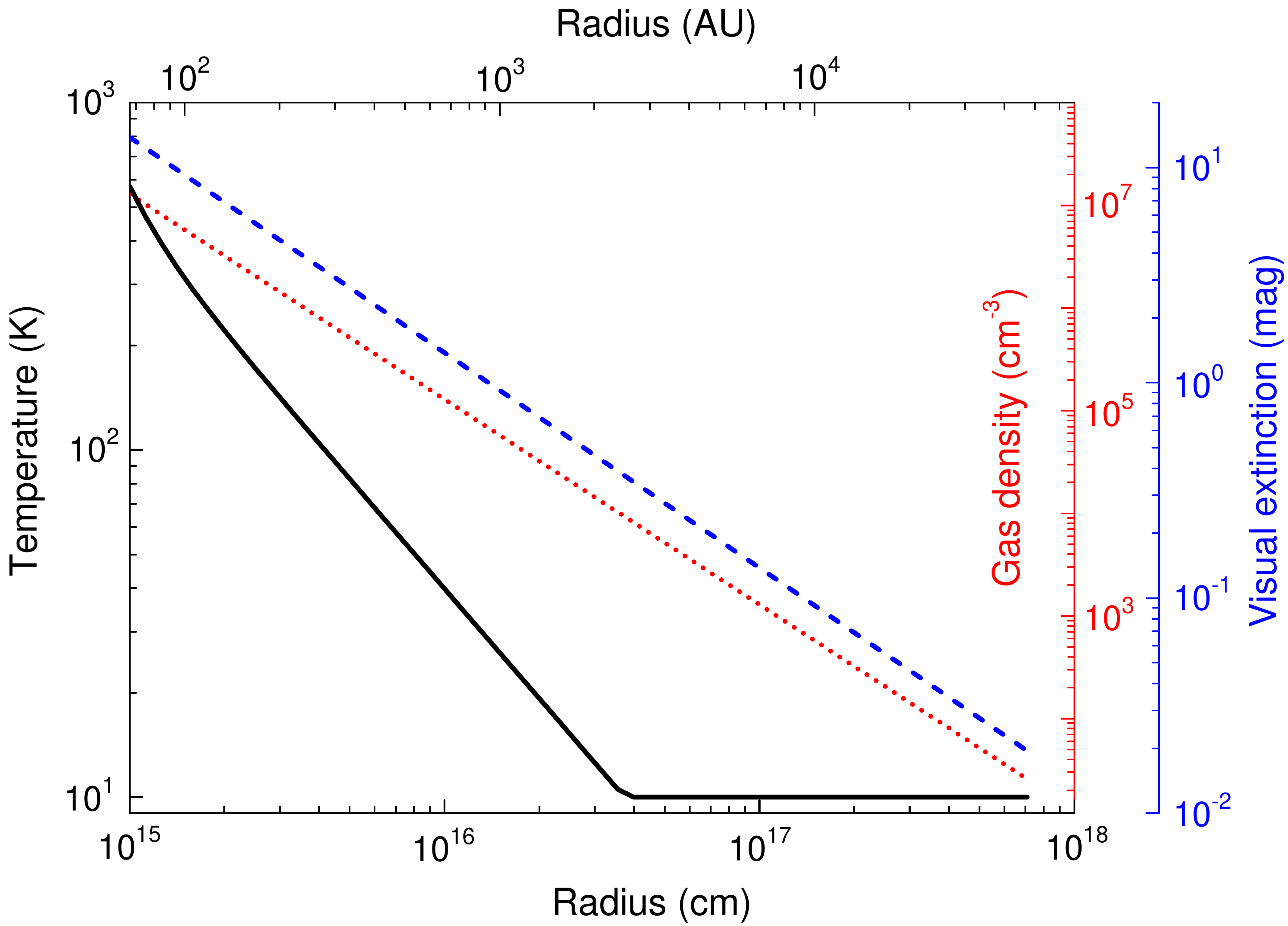}
\caption{Gas density, visual extinction, and gas temperature as functions of radius for the CSE of IRC +10216 from the center of the star towards the outside of the envelope. For clarity, the radius is given in units of `cm' and `AU'.   }
\label{figT}  
\end{figure}

}

\def\placefigureaDvscD
{  

\begin{figure}
\includegraphics[width=0.46\textwidth]{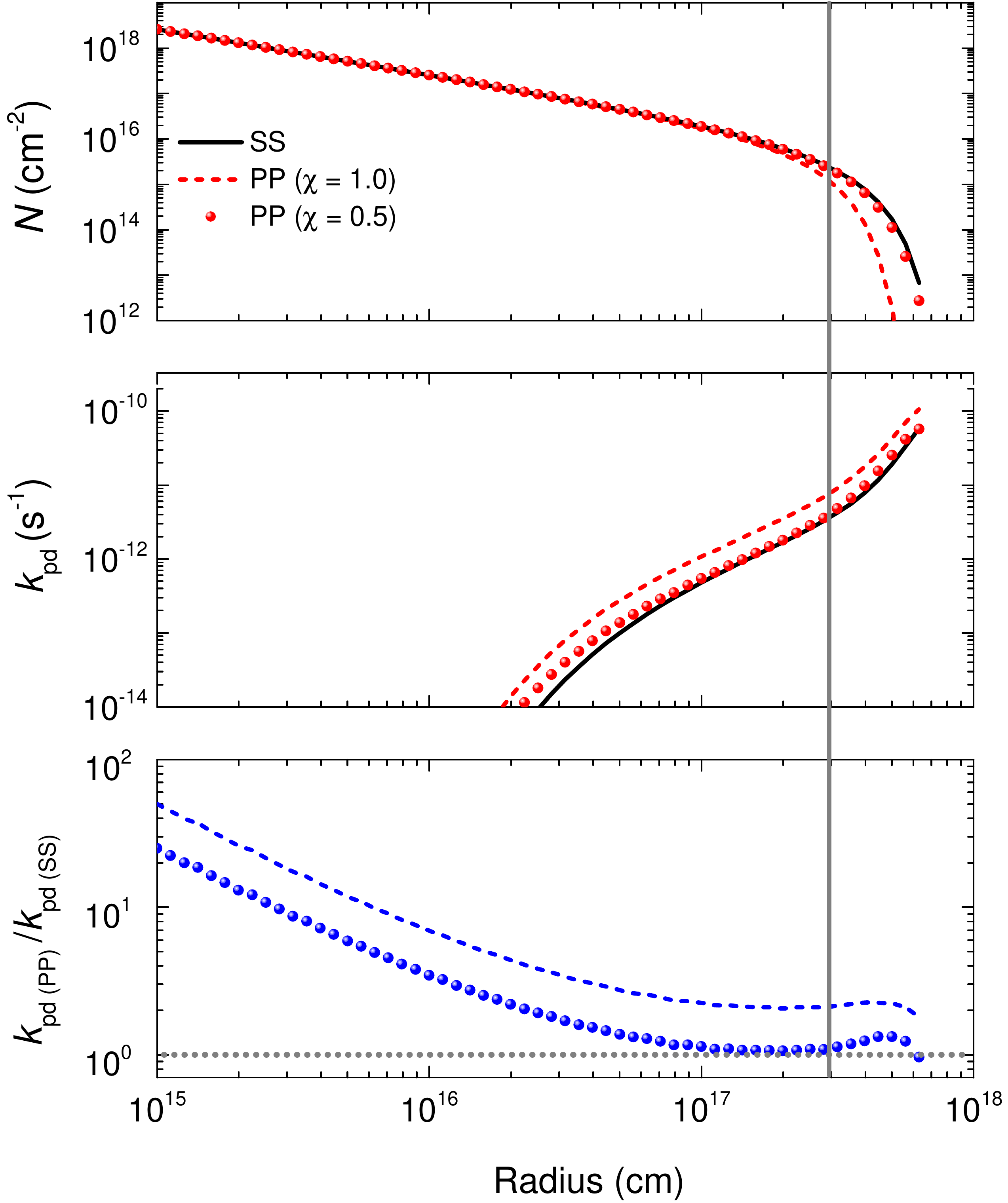}
\caption{\textcolor{black}{Plots of the \ce{N2} column densities (top), photodissociation rates (middle) and their ratios (bottom) when calculated using the plane-parallel (dashed lines: $\chi=1.0$; spheres: $\chi=0.5$) and spherically-symmetric (black solid lines) models as a function of radius. The dotted horizontal line indicates $k_{\rm {pd}}(\rm {PP})$ = $k_{\rm{pd}}(\rm {SS})$. Outside the solid vertical line \ce{N2} photodissociation becomes considerable.}   \label{figaDvscD}}
\end{figure}

}

\def\placefigureNb
{  

\begin{figure}
\includegraphics[angle=0,width=0.45\textwidth]{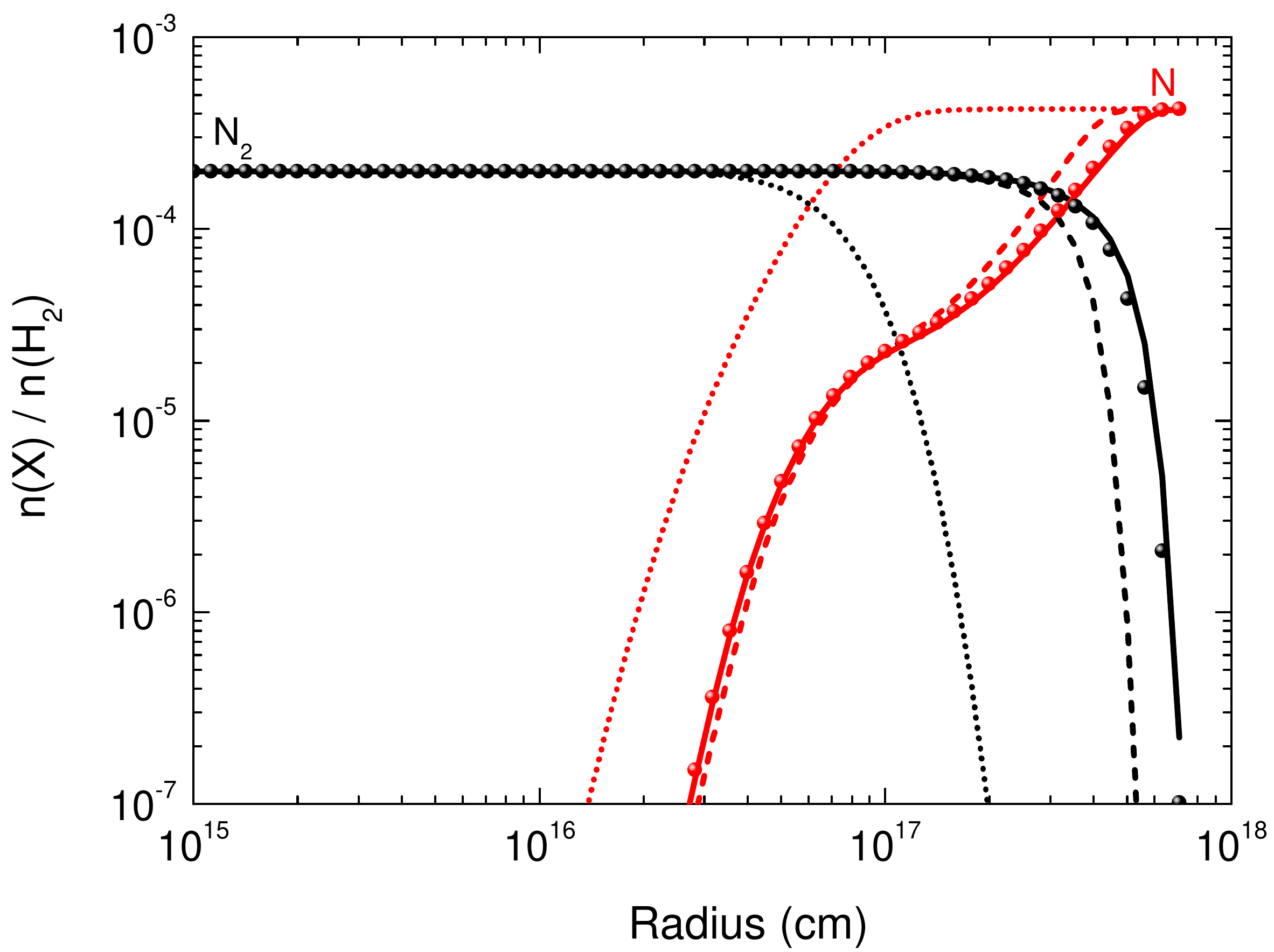}
\caption{Plot of the fractional abundances, relative to \ce{H2}, of N$_2$ and N as a function of radius. Dotted lines: \citet{McElroy13} model, dust shielding only. \textcolor{black}{ Spheres, dashed and solid lines: PP model ($\chi=0.5$), PP model ($\chi=1.0$), and SS model, with full shielding (dust + self- + H + H$_2$). }  }

\label{figNb}
\end{figure}

}

\def\placefigurediffshielding
{  

\begin{figure}
\includegraphics[angle=0,width=0.45\textwidth]{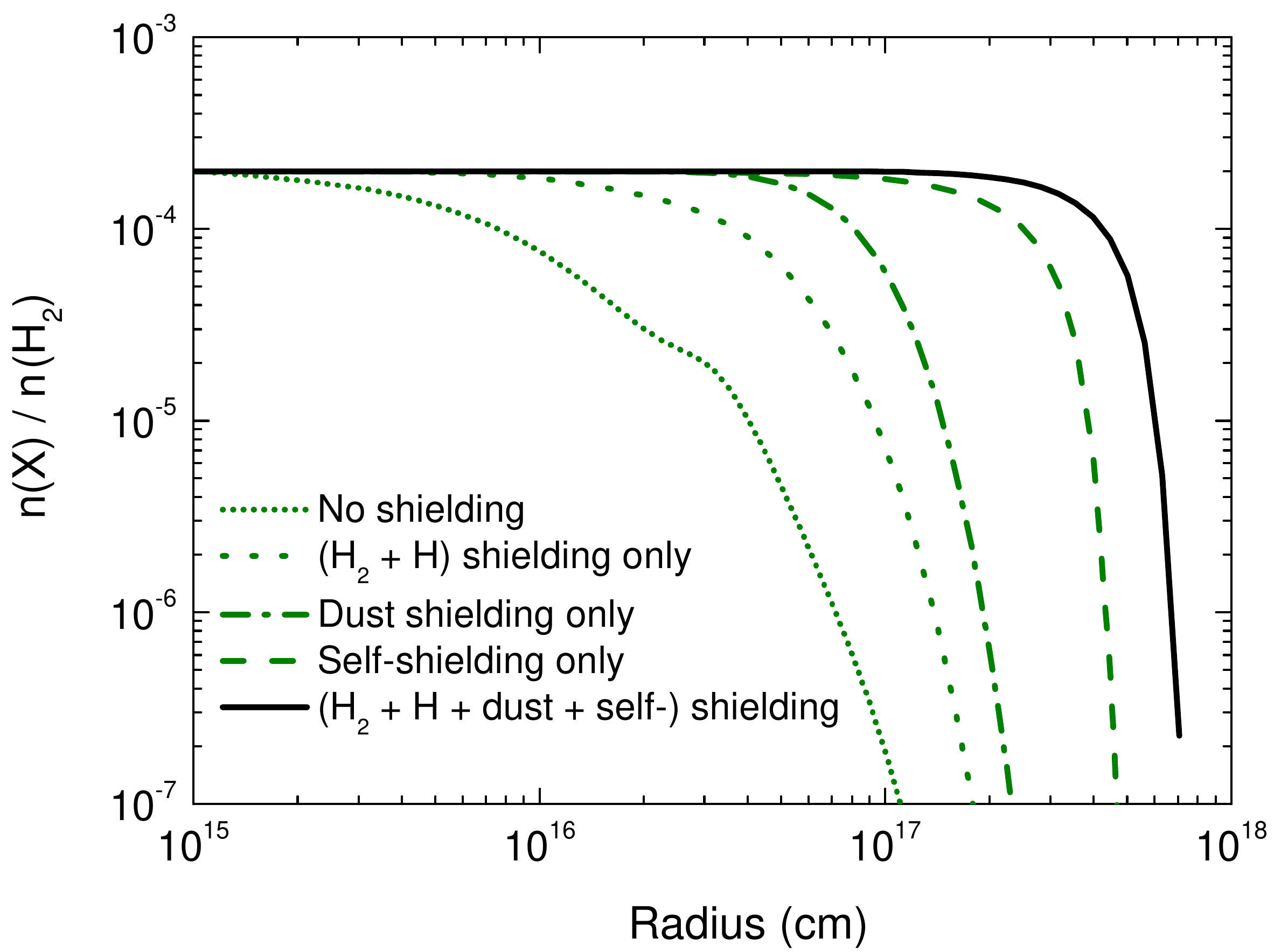}
\caption{Plot of the fractional abundance of N$_2$, relative to \ce{H2}, as a function of radius with different shielding effects. In all cases, photons are considered from all directions in space (in SS model).           } 
\label{figdiffshielding}
\end{figure}

}

\def\placefigureCO
{  

\begin{figure}
\subfigure{\includegraphics[width=0.45\textwidth]{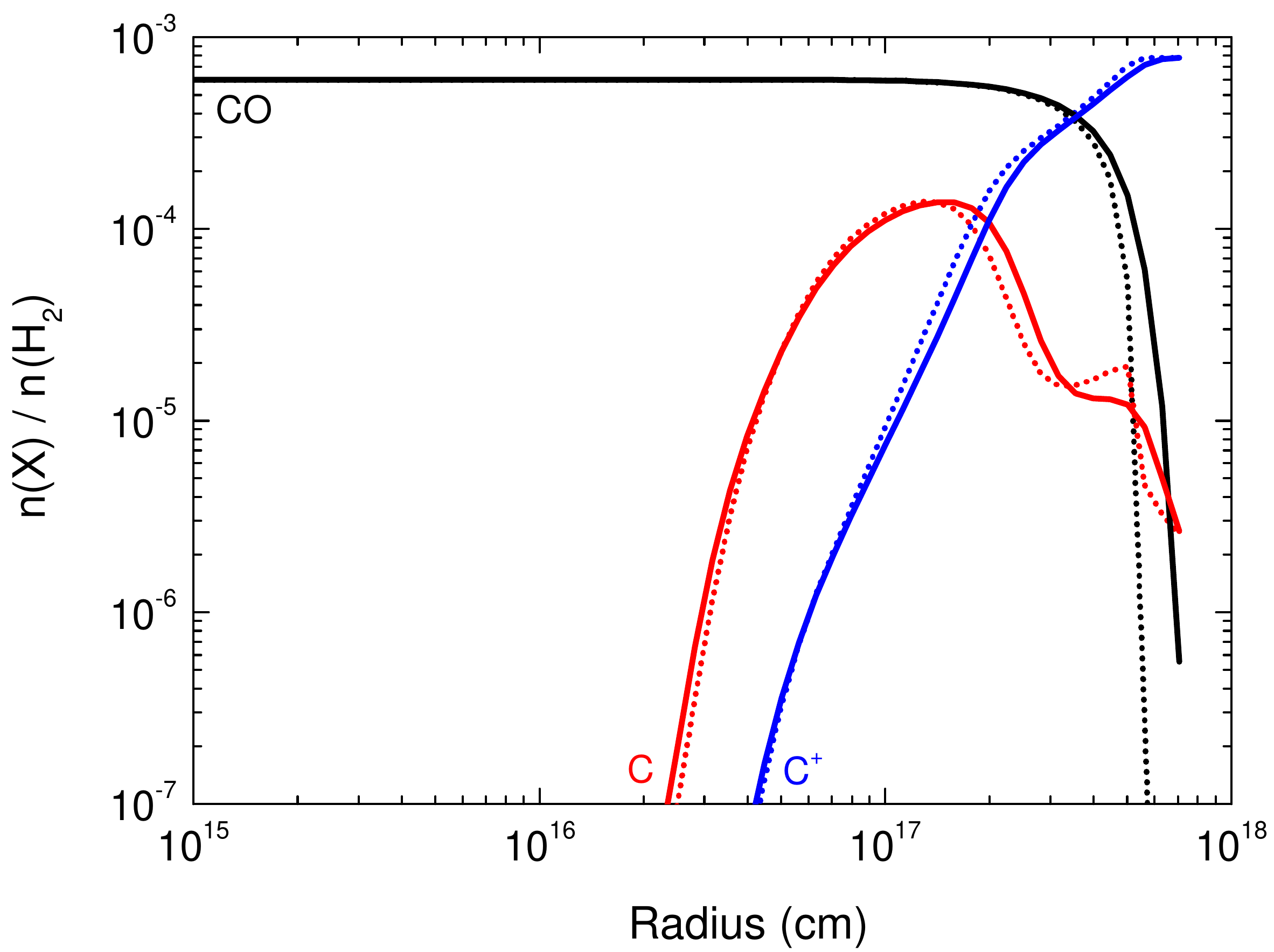}}
\caption{Plot of the fractional abundances of CO, C, and C$^+$, relative to \ce{H2}, as a function of radius. Dotted lines: \citet{McElroy13} model employing an earlier CO self-shielding function \citep{Morris83} and photodissociation rates \citep{vanDishoeck88}. \textcolor{black}{Solid lines: SS model,} employing the newly calculated self-shielding function and photodissociation rate of CO \citep{Visser09}.    }

\label{figCO}
\end{figure}

}

\def\placefigureCaceN
{

\begin{figure}
\includegraphics[width=0.45\textwidth]{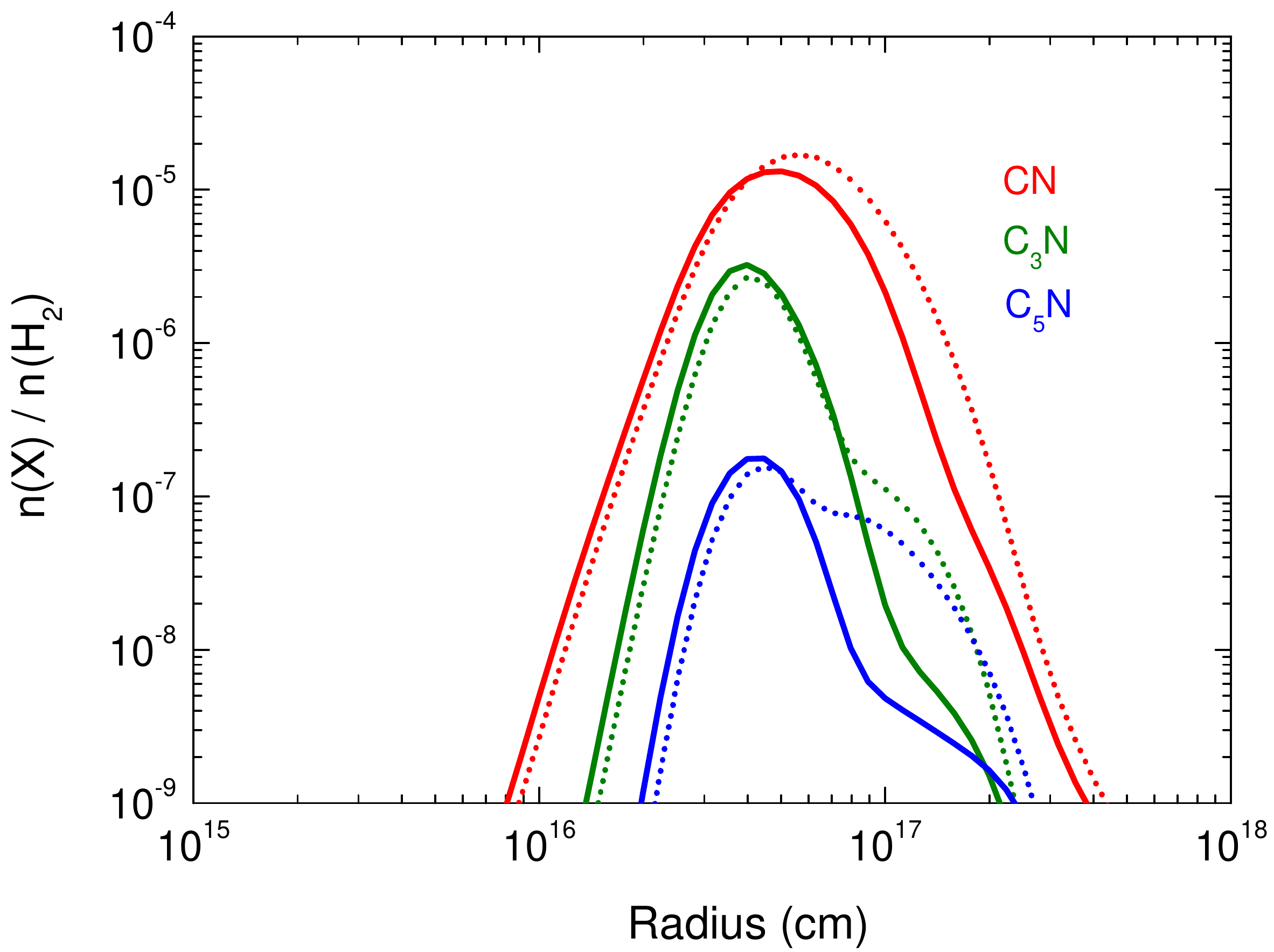}
\caption{Plot of the fractional abundances, relative to \ce{H2}, of C$_n$N ($n$=1, 3, 5) as functions of radii. Dotted and solid lines exhibit the results from the model of \citet{McElroy13} and the SS model. \label{figCaceN}}
\end{figure}

}

\def\placefigureCaceNy
{

\begin{figure}
\includegraphics[width=0.45\textwidth]{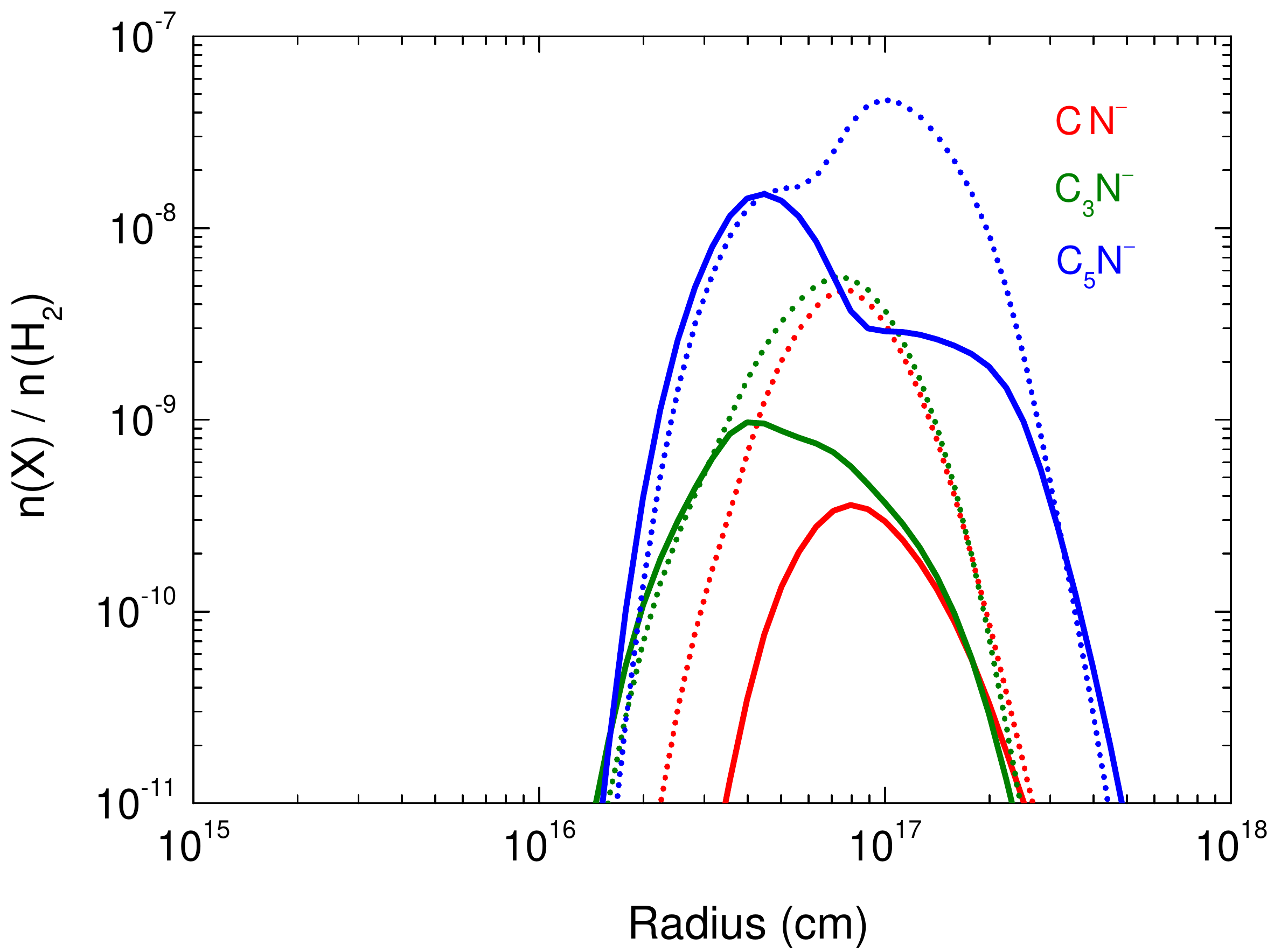}
\caption{Plot of the fractional abundances, relative to \ce{H2}, of C$_n$N$^-$ ($n$=1, 3, 5) as a function of radius. Dotted lines: \citet{McElroy13} model. Solid lines: SS model with updated shielding of \ce{N2} and CO, details are described in Section \ref{subsection:What's new}.    \label{figCaceNy}}
\end{figure}

}

\def\placefigureCgiN
{  

\begin{figure}
\includegraphics[width=0.45\textwidth]{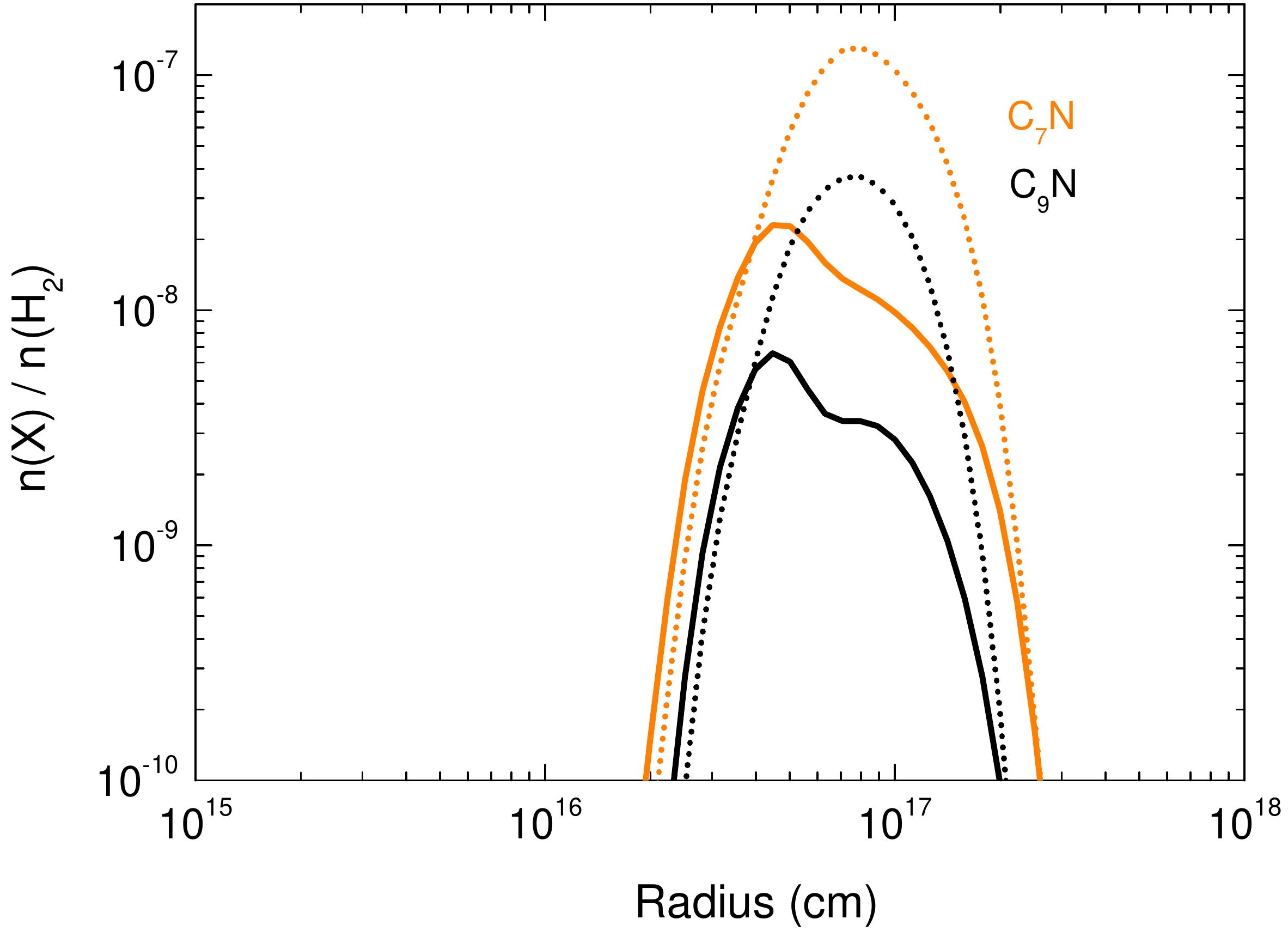}
\caption{Plot of the fractional abundances, relative to \ce{H2}, of C$_7$N and C$_9$N as a function of radius from the center of the star towards the outside of the envelope.  Dotted and solid lines exhibit the results from the model of \citet{McElroy13} and the SS model. \label{figCgiN}}
\end{figure}

}

\def\placefigureHCacegiN
{

\begin{figure}
\includegraphics[width=0.46\textwidth]{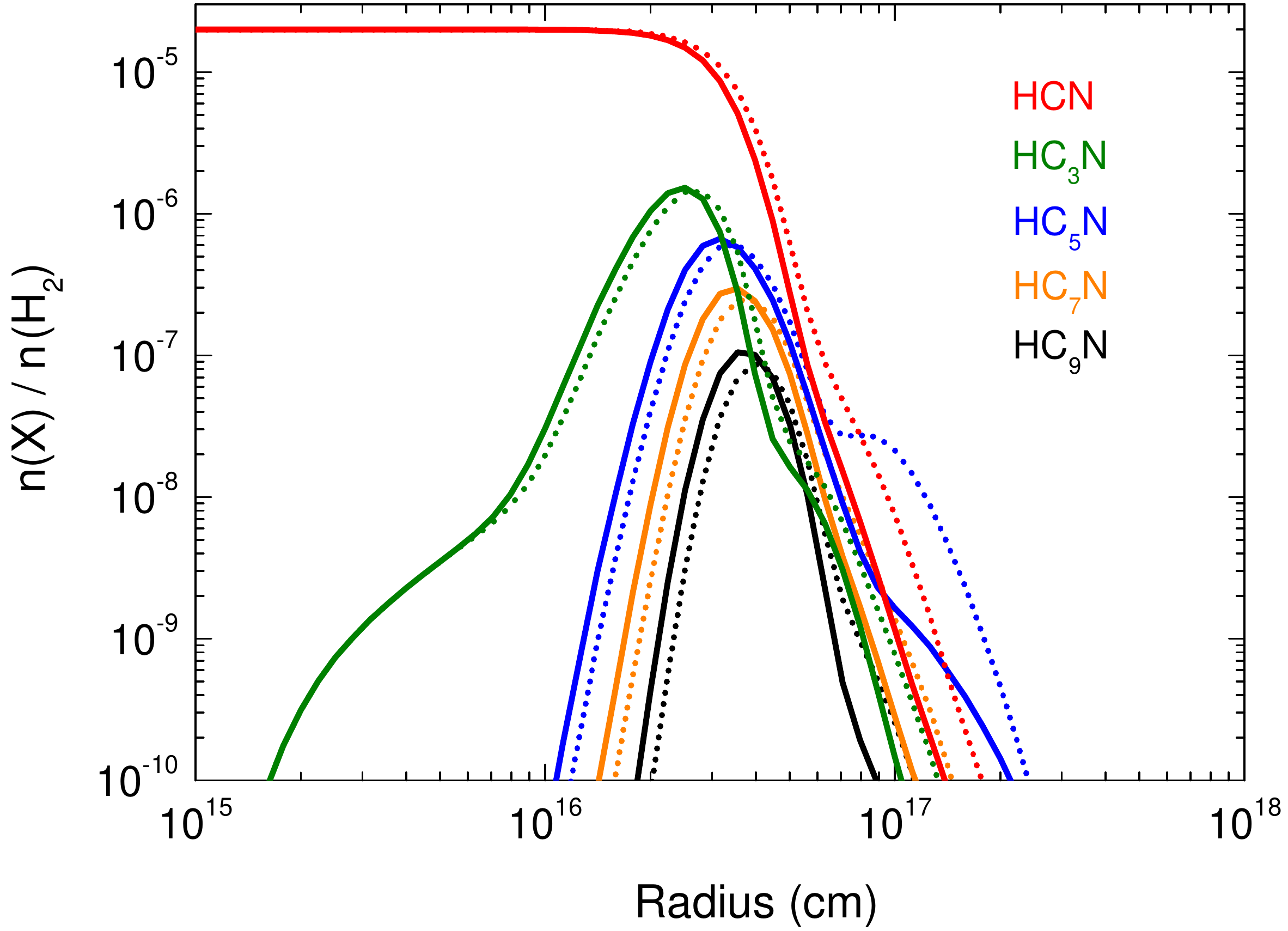}
\caption{Plot of the fractional abundances, relative to \ce{H2}, of HC$_n$N ($n$=1, 3, 5, 7, 9) as a function of radius from the center of the star towards the outside of the envelope.  Dotted and solid lines exhibit the results from the model of \citet{McElroy13} and the SS model. 
\label{figHCacegiN}}
\end{figure}

}

\def\placefigureHbCN
{  
\begin{figure}
\vspace{-1.6ex}  
\includegraphics[width=0.46\textwidth]{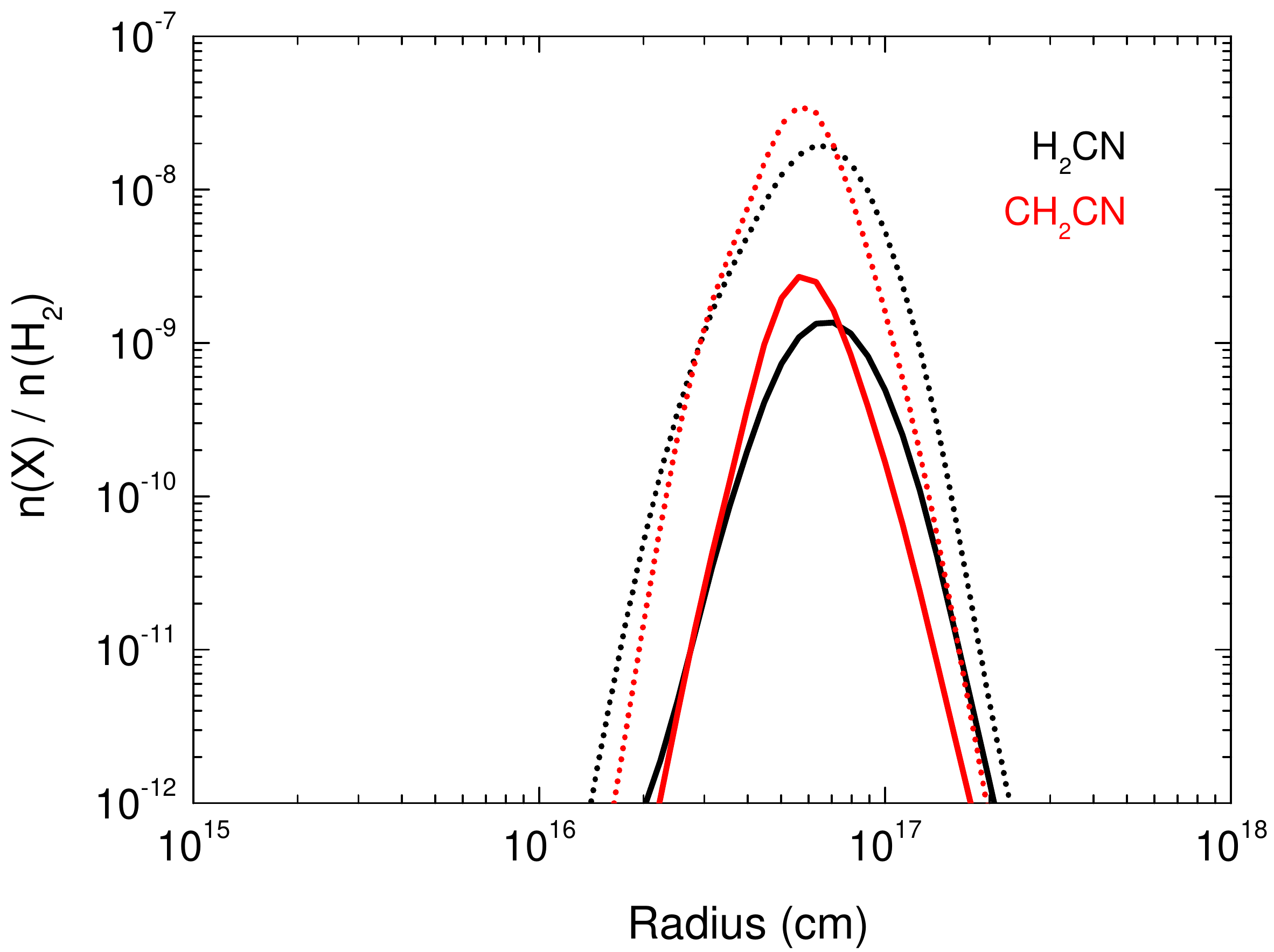}
\caption{Plot of the fractional abundances, relative to \ce{H2}, of H$_2$CN and CH$_2$CN as a function of radius from the center of the star towards the outside of the envelope.   Dotted and solid lines exhibit the results from the model of \citet{McElroy13} and the SS model.  \label{figHbCN}}
\end{figure}

}

\def\placefigurePN
{

\begin{figure}
\includegraphics[width=0.45\textwidth]{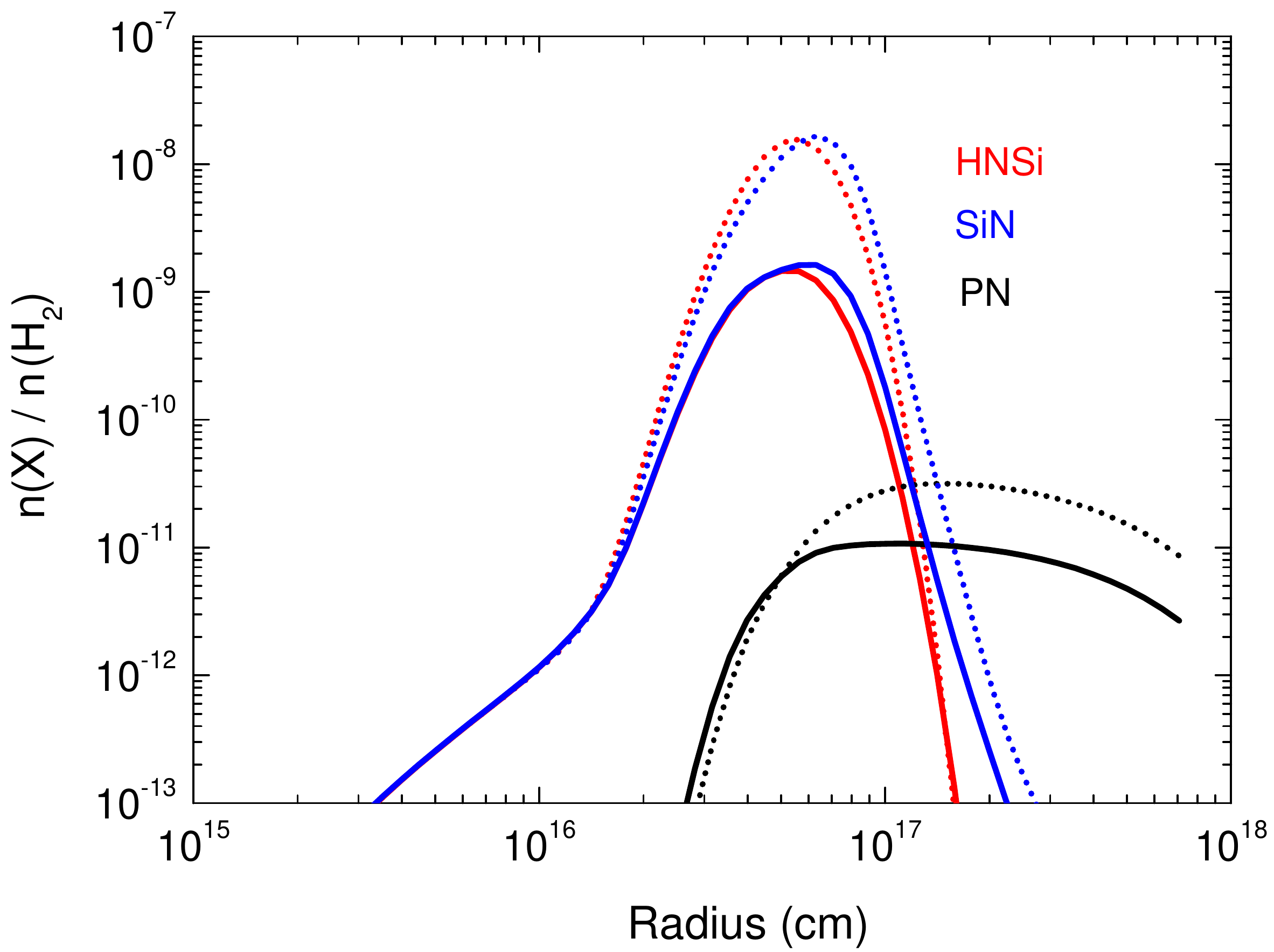}
\caption{Plot of the fractional abundances, relative to \ce{H2}, of PN, SiN and HNSi as a function of radius from the center of the star towards the outside of the envelope.   Dotted and solid lines exhibit the results from the model of \citet{McElroy13} and the SS model.  \label{figPN}}
\end{figure}

}

\def\placefigureNHc
{  
\begin{figure}
\includegraphics[width=0.45\textwidth]{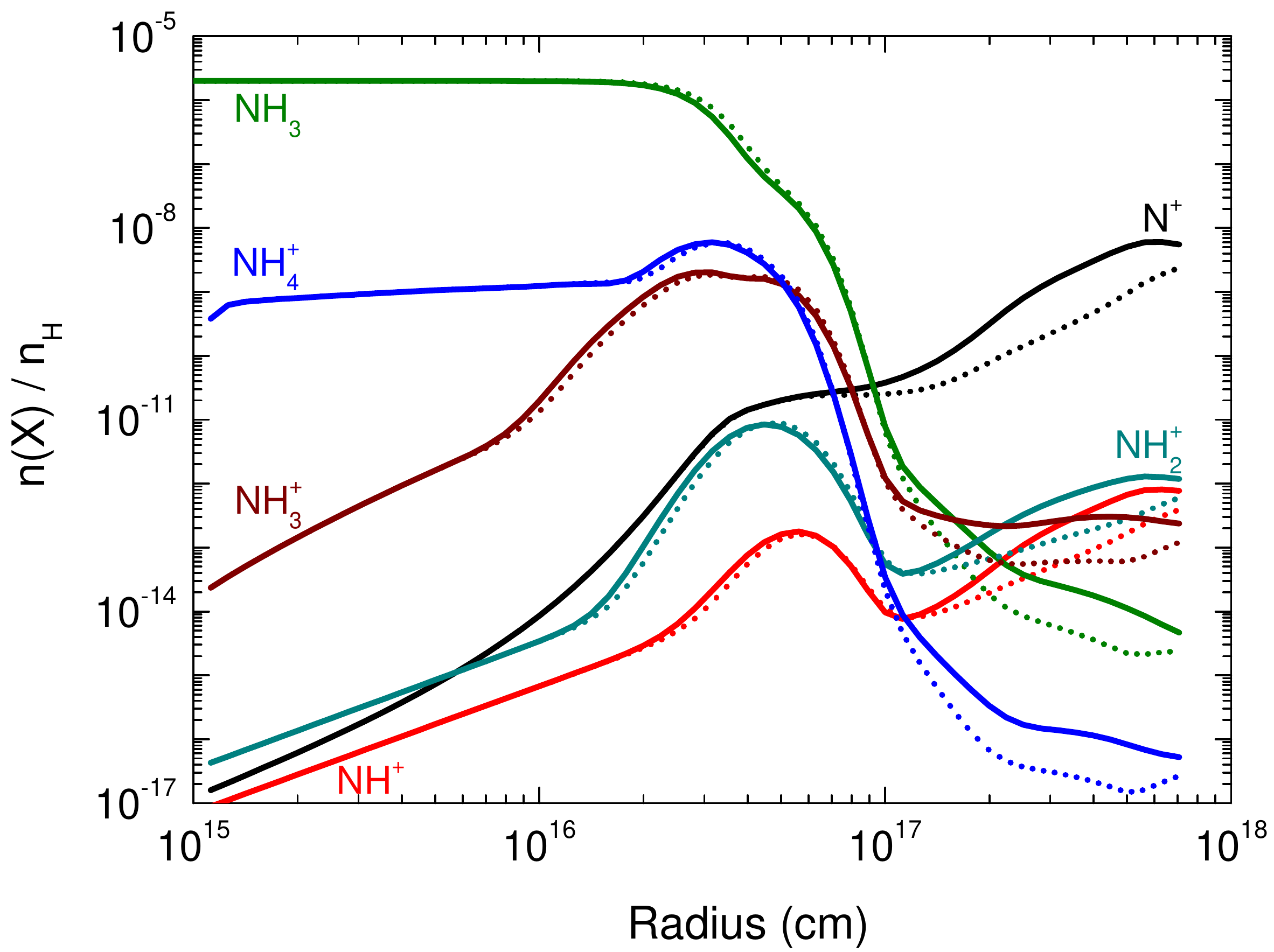}
\caption{Plot of the fractional abundances, relative to \ce{H2}, of N$^+$, NH$^+$, NH$_2^+$, NH$_3^+$, NH$_4^+$ and NH$_3$ as a function of radius from the center of the star towards the outside of the envelope.   Dotted and solid lines exhibit the results from the model of \citet{McElroy13} and the SS model.  \label{figNHc}}
\end{figure}

}

\def\placefigureNbHx
{  

\begin{figure}
\includegraphics[width=0.45\textwidth]{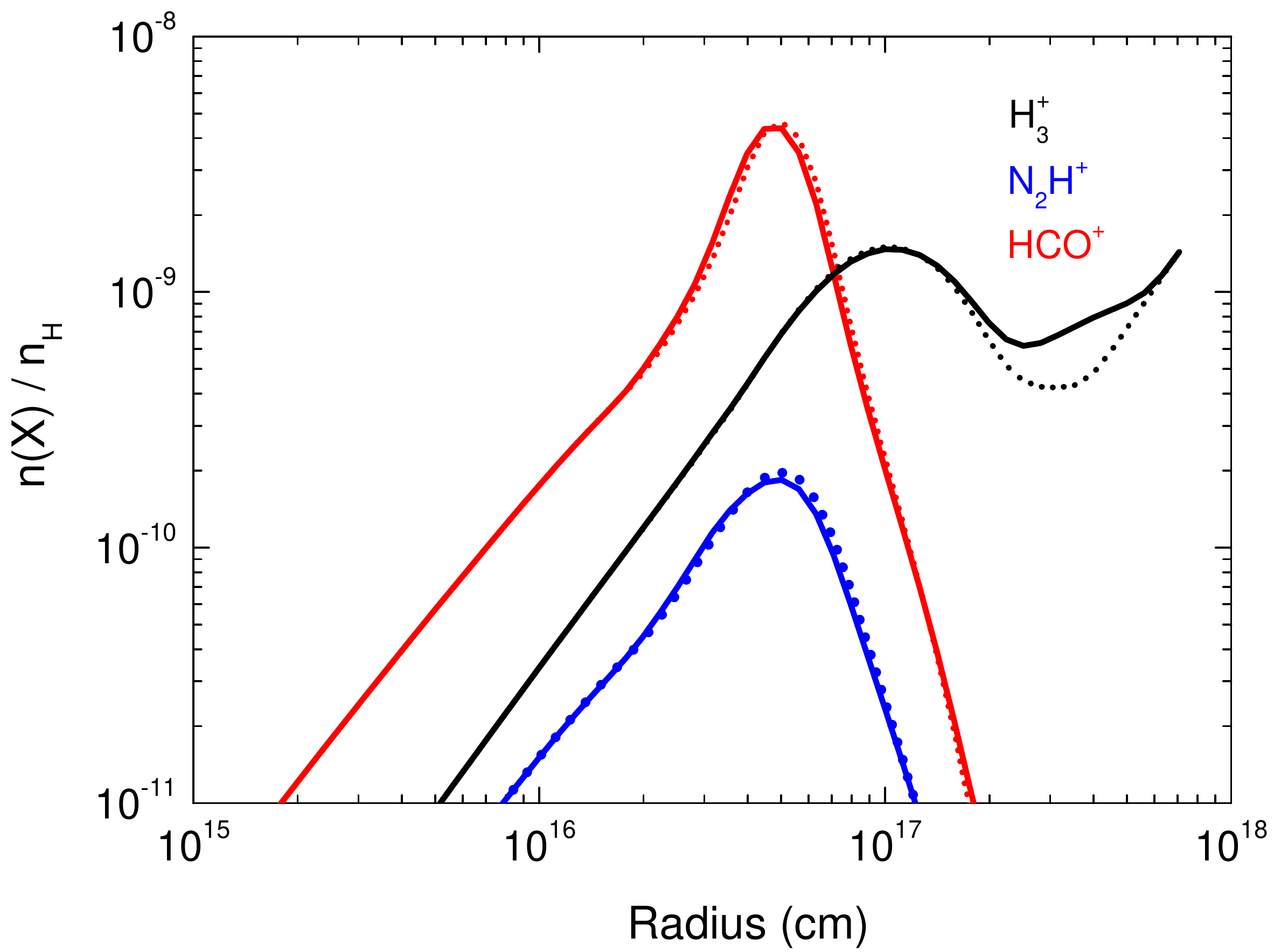} 
\caption{Plot of the fractional abundances, relative to \ce{H2}, of HCO$^+$ and N$_2$H$^+$ as a function of radius from the center of the star towards the outside of the envelope. Dotted and solid lines exhibit the results from the model of \citet{McElroy13} and the SS model. \label{figNbHx}}
\end{figure}

}

\def\placefigurepeakradius
{  

\begin{figure}
\includegraphics[width=0.4\textwidth]{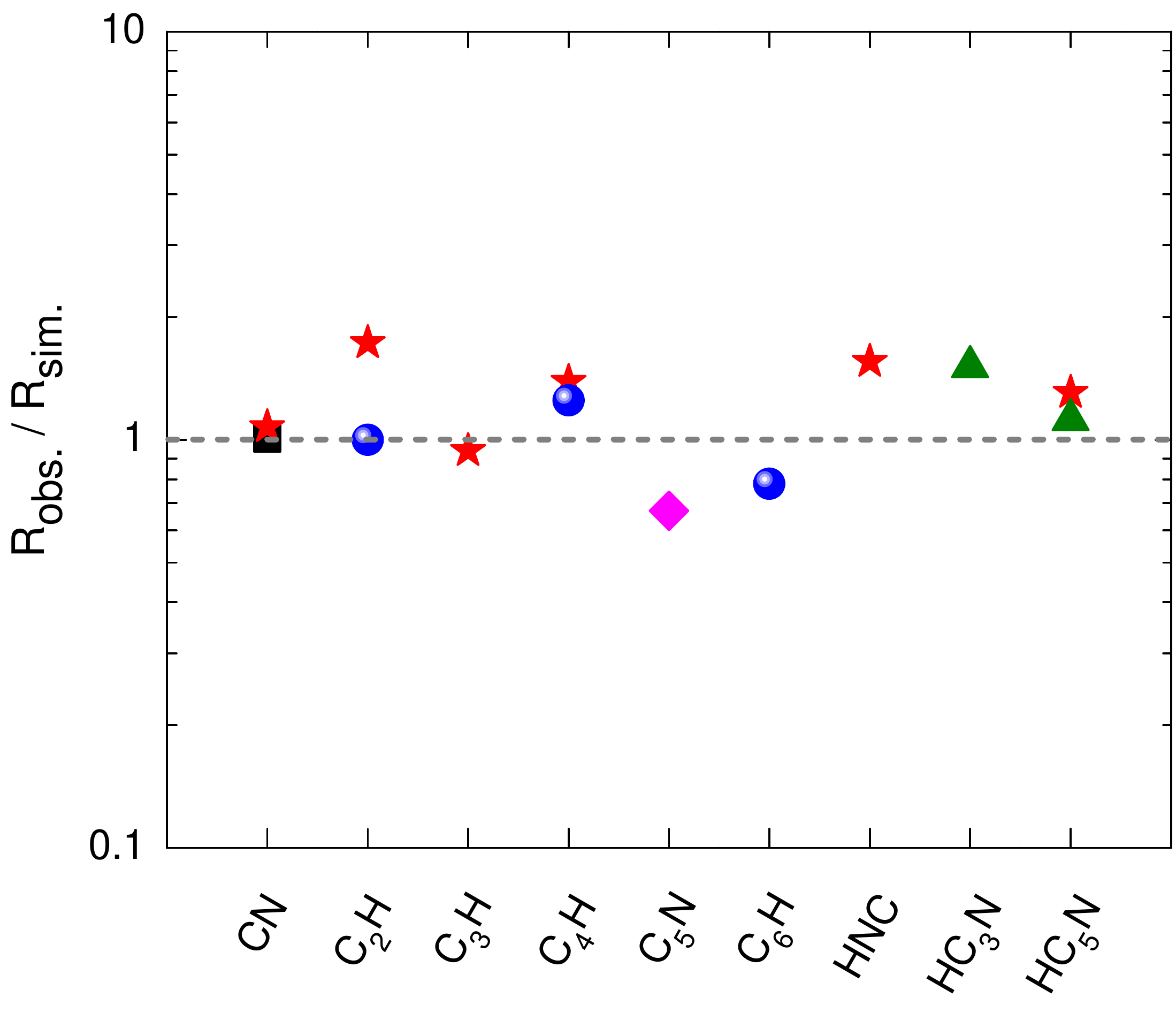}
\caption{Ratio of observed and simulated peak abundance radii ($R_{\rm obs.}$ and $R_{\rm sim.}$) for selected species. 
In all cases, $R_{\rm sim.}$ is obtained from our SS model.
$R_{\rm obs.}$ is estimated from the interferometric maps of \citet{Lucas95} (black squares), \citet{Guelin98} (pink diamond), \citet{Guelin99} (blue spheres), \citet{Dinh-V-Trung08} (green triangles), and \citet{Guelin11} (red stars). The dashed horizontal line indicates the position where  $R_{\rm sim.}$ $=$ $R_{\rm obs.}$.   
\label{figpeakradius}}

\end{figure}

}

\def\placefigurediffNb
{  

\begin{figure}
\includegraphics[width=0.5\textwidth]{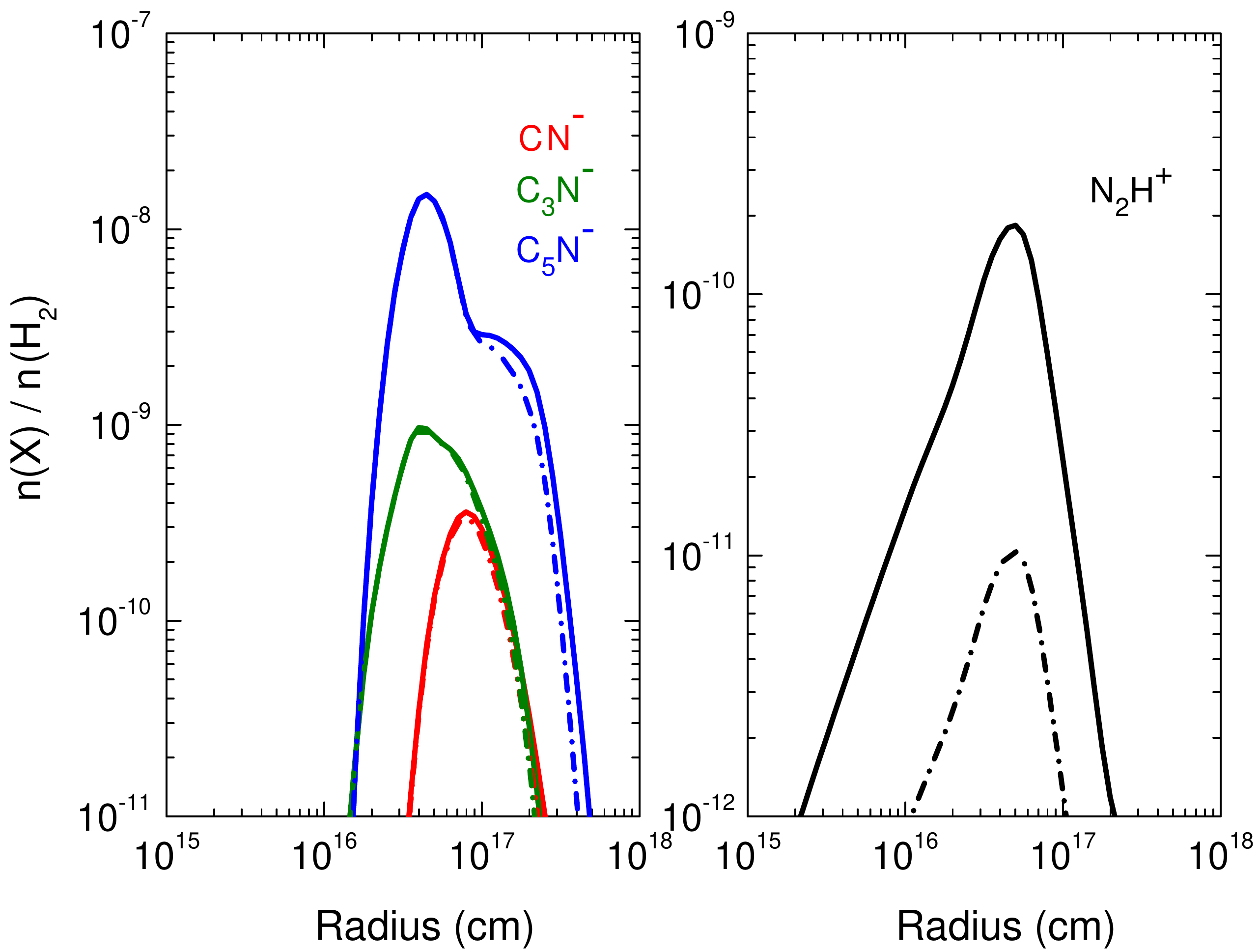} 
\caption{Plot of the fractional abundances, relative to \ce{H2}, of \ce{C_nN^-} (n = 1, 3, 5) (left panel) and \ce{N2H+} (right panel) as a function of radius. Solid lines: calculated  \textcolor{black}{using} an initial \ce{N2} abundance of $2 \times 10^{-4}$ (relative to \ce{H2}). Dash-dotted lines: calculated  \textcolor{black}{using} an initial \ce{N2} abundance of $1 \times 10^{-5}$. In both cases, the SS model with updated shielding was employed.  \label{figdiffNb}}
\end{figure}

}

\def\placefigurecDmodel
{  

\begin{figure}
\includegraphics[width=0.5\textwidth]{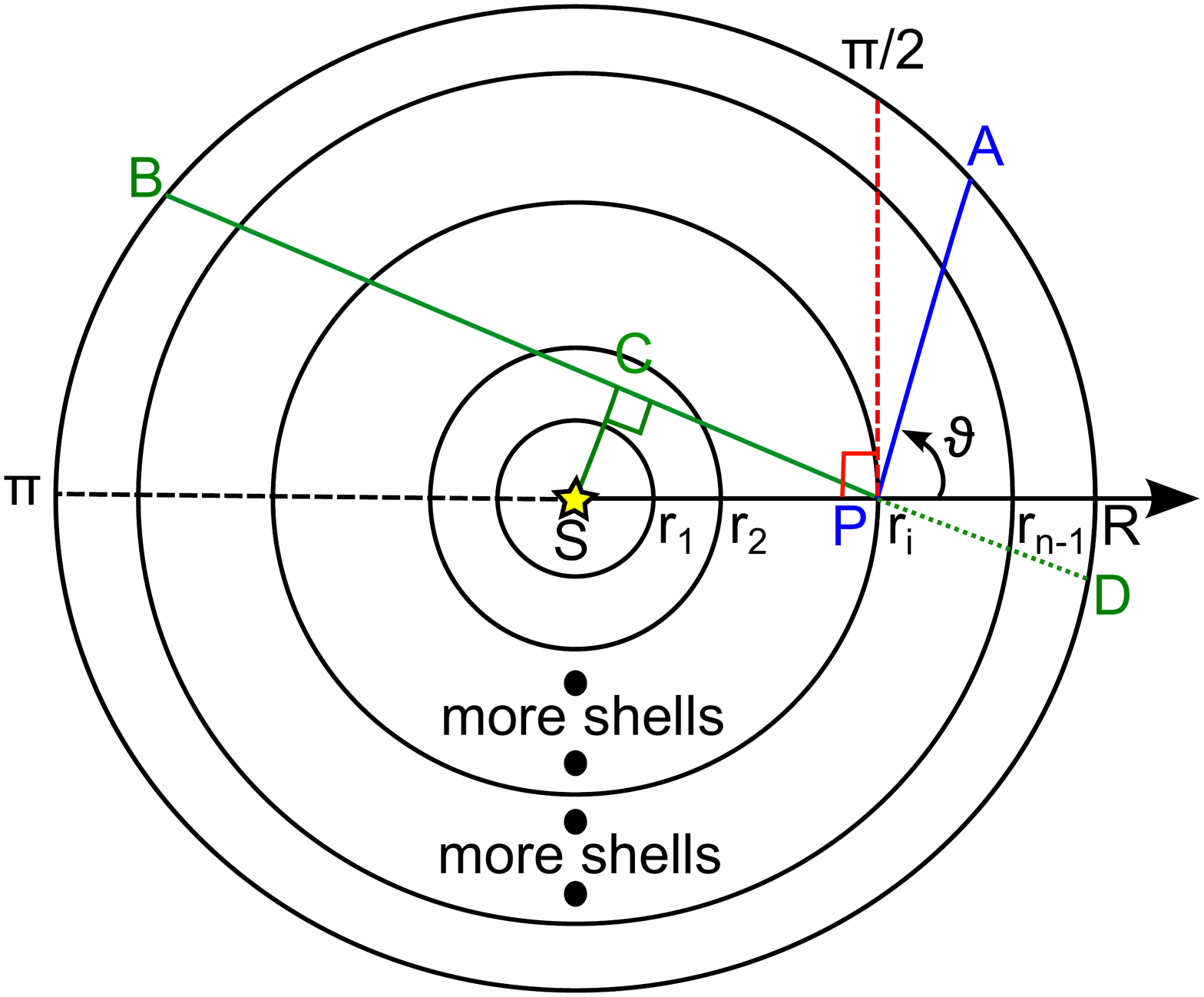}
\caption{Structure of the circumstellar envelope model of an AGB star.  \label{fig3D_model}}

\end{figure}

}

\def\placefigureNbrceta
{  

\begin{figure}
\includegraphics[angle=0, width=0.5\textwidth]{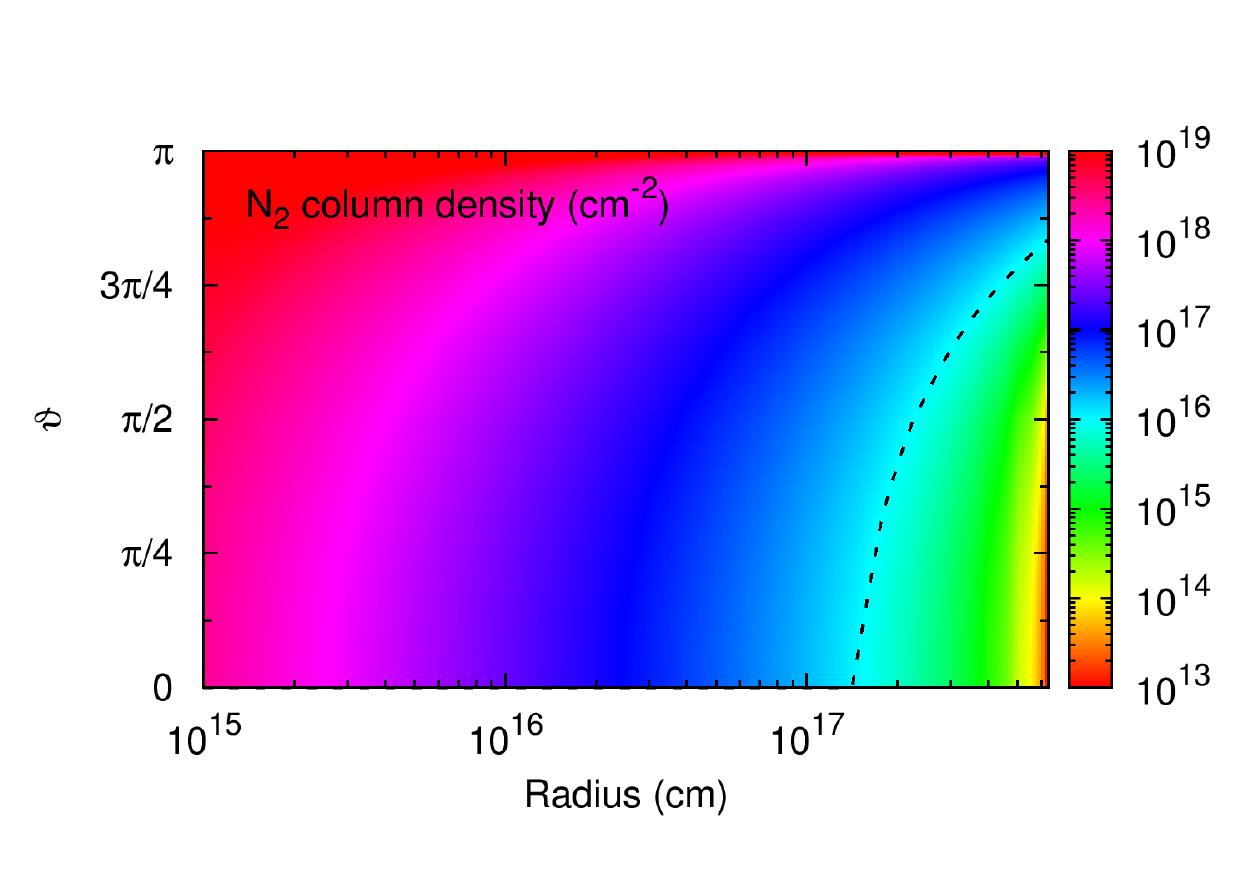}
\caption{Map of N$_2$ column density as a function of radius and the angle $\vartheta$ with respect to the normal direction, as described in Fig. \ref{fig3D_model}. The dashed line shows the critical column density  (1.0 $\times$ $10^{16}$ cm$^{-2}$). In practice, molecular N$_2$ is $\sim$ fully shielded when its column density is higher than this value.  \label{figNbrceta}}
\end{figure}

}




\def\placetabIrcParameters
{
\begin{table*}
  \caption{Envelope parameters and assumptions for IRC +10216 in this study.\tablefootmark{a} }
\label{tab:EnvelopeParameters}
\centering
\resizebox{0.8\linewidth}{!}{ 
\begin{tabular}{lcl}
\hline\hline
 ~~1. Shape                           &       & Spherical                       \\
 ~~2. Mass-loss rate                  &       & 1.5(-5) M$_\odot$ yr$^{-1}$      \\ 
 ~~3. Envelope expansion velocity      &       & 14.5 km s$^{-1}$                 \\ 
 ~~4. Radiation field                 &       & Standard interstellar radiation field \citep{Draine78}, isotropic incidence \\
 ~~5. Dust and gas shells in the outflow &    & Ignored                                                  \\
 ~~6. Grain surface reactions          &        & Ignored                         \\
 ~~7. Polycyclic aromatic hydrocarbons (PAHs)   &      &  Ignored \\
 ~~8. Gas density distribution        &        & Falls as $r^{-2}$, where $r$ is the distance from the central star \\ 
 ~~9. Chemical evolution               &        & Kinetic equations solved as a function of radius as material traverses the CSE \\
   10. H$_2$                           &        & Fully self-shielded, no photodissociation                   \\
   11. Parent species                 &        & See Table~\ref{tab:InitialABofParentSpecies}         \\
   12. $T$,  $A_{\rm V}$ , and gas density     &        & See Fig.~\ref{figT}                                 \\        
   13. Distance  &   & 150 pc \citep{De_Beck12}           \\
\hline
\end{tabular}
} 
\tablefoot{
\tablefoottext{a}{a(b)=$a\times 10^{b}$}. } 
\end{table*}
}


\def\placetabInitialAB
{
\begin{table}
  \caption{ Initial abundances of parent species, relative to \textcolor{black}{\ce{H2}}, at the inner radius.\tablefootmark{a} }
    
\label{tab:InitialABofParentSpecies}
\centering
\resizebox{0.68\linewidth}{!}{ 

\begin{tabular}{lclc}
\hline\hline
Species & Abundance   & Species & Abundance  \\
\hline
 N$_2$  &   2.0(-4)   & SiS     & 1.3(-6)    \\ 
 NH$_3$ &   2.0(-6)   & CH$_4$  & 3.5(-6)    \\ 
 HCN    &   2.0(-5)   & H$_2$O  & 1.0(-7)    \\ 
 He     &   1.0(-1)   & Mg      & 1.0(-5)    \\ 
 HF     &   8.0(-9)   & C$_2$H$_4$ & 2.0(-8) \\ 
 C$_2$H$_2$ & 8.0(-5) & SiH$_4$ & 2.2(-7)    \\ 
 CO     &   6.0(-4)   & HCl     & 1.0(-7)    \\ 
 H$_2$S &   4.0(-9)   & HCP     & 2.5(-8)    \\ 
 CS     &   7.0(-7)   & SiC$_2$ & 2.0(-7)    \\ 
 SiO    &   1.8(-7)   &         &            \\ 

\hline

\end{tabular}

} 
\tablefoot{
\tablefoottext{a}{Same as those adopted in the model of \citet{McElroy13}. 
} 
}
\end{table}
}

%

\def\placetabNcol
{
\begin{table*}
\caption{Calculated and observed column densities, as well as peak abundance radii of species detected in the CSE of IRC +10216 \tablefootmark{a}.} \label{tab:Ncol}
\centering
\resizebox{0.9\linewidth}{!}{ 
\begin{tabular}{llcccccc}
\hline\hline	
No. & Species \tablefootmark{b} 	&	Observed\tablefootmark{c}	&	Calculated\tablefootmark{d}	&	Calculated \tablefootmark{e} 	&	Percentage change\tablefootmark{f} 	&	Radius of peak\tablefootmark{g}  &  Radius of peak\tablefootmark{h}				\\
	&			&				&	(This work)	&	(R{\tiny ATE}12 model)	&	between the     &	abundance &  emission        \\
	&           &  (cm$^{-2}$)  &   (cm$^{-2}$) &   (cm$^{-2}$) &  two simulations   &   (cm \& arcsecond, $^{\prime \prime }$) &(cm \& arcsecond, $^{\prime \prime }$)                      \\
\hline						
1	&	CN		&	1.1(15)	&	3.5(15)	&	4.1(15) 	&												 	-15	&	5.0(16)	\& 22	& 5.4(16) \& 24		\\
2	&	\textbf{C$_3$N}	&\textbf{	2-4(14)}	&\textbf{	6.7(14)}	&\textbf{	5.1(14)} 	&			\textbf{		31}	&	4.0(16)	\& 18			\\
3	&	C$_5$N	&	3-6(12)	&	3.6(13)	&	3.5(13) 	&													3	&	4.5(16)	\& 20		& 3.0(16) \& 13	\\
4	&	\textbf{CN$^{-}$}     	&	\textbf{5.0(12)}	&	\textbf{6.0(10)}	&	\textbf{7.0(11)}   	&\textbf{	-91}	&	7.9(16)	\& 35			\\
5	&	\textbf{C$_3$N$^{-}$} 	&	\textbf{2.0(12)}	&	\textbf{3.2(11)}  	&	\textbf{1.0(12)}   	&	\textbf{-68}	&	4.0(16)	\& 18			\\
6	&	\textbf{C$_5$N$^{-}$} 	&	\textbf{3.0(12)}	&	\textbf{3.8(12)}	&	\textbf{7.7(12)} 	&	\textbf{-51}	&	4.5(16)	\& 20			\\
7	&	HC$_3$N	&	1-2(15)	&	5.4(14)	&	4.8(14) 	&													13	&	2.5(16)	\& 11		 & 3.8(16) \& 17	\\
8	&	HC$_5$N	&	2-3(14)	&	1.7(14)	&	1.4(14)	&														21	&	3.2(16)	\& 14			\\
9	&	HC$_7$N	&	1.0(14)  	&	6.0(13)	&	4.7(13) 	&												28	&	3.5(16)	\& 16			\\
10	&	\textbf{HC$_9$N}	&	\textbf{3.0(13)}  	&	\textbf{1.8(13)}	&	\textbf{1.3(13)} 	&												\textbf{38}	&	3.5(16)	\& 16   & 4.3(16) \& 19 			\\
11	&	\textbf{PN}   	&	\textbf{1.0(13)}	&	\textbf{2.6(09)} 	&	\textbf{5.4(09)}	&\textbf{		-52}	&	\textbf{1.1(16)}	\& ~ \textbf{5}			\\
12	&	\textbf{SiN}  	&	\textbf{4.0(13)}	&	\textbf{3.5(11)}	&	\textbf{2.4(12)} 			&	\textbf{-85}	&	6.3(16)	\& 28			\\
13	&	SiNC	&	2.0(12)	&	8.7(08) 	&	1.2(09)     		&	-28	&	4.5(16)	\& 20			\\
14	&	CH$_2$NH  	&	9.0(12) 	&	2.8(11)	&	2.4(11) 											&	17	&	2.8(16)	\& 12			\\
15	&	\textbf{CH$_2$CN}  	&	\textbf{8.4(12)} 	&	\textbf{3.0(11)}	&	\textbf{4.2(12)} 		&	\textbf{-93}	&	5.6(16)	\& 25			\\
16	&	CH$_2$CHCN	&	5.0(12) 	&	5.1(11)	&	5.3(11) 											&	-4	&	5.6(16)	\& 25			\\
17	&	CH$_3$CN  	&	6-30(12)	&	4.9(12)	&	5.6(12) 		&	-13	&	4.0(16)		\& 18		\\
18	&	C                	&	1.1(16)	&	1.9(16)	&	1.9(16)	&											0	&	1.4(17)		\& 62		\\
19	&	C$_2$      	&	 7.9(14)  	&	5.2(15)  	&	4.2(15) 	&										24	&	5.0(16)	\& 22			\\
20	&	C$_3$       	&	 1.0(15)    	&	2.0(14) 	&	1.8(14) 	&									11&	3.5(16)	\& 16			\\
21	&	\textbf{C$_5$}      	&	 1.0(14)    	&	2.0(14) 	&	1.5(14) 	&										\textbf{33}&	7.1(16) \& 32				\\
22	&	CP         	&	 1.0(14)    	&	2.4(12)	&	2.2(12) 	&										9	&	4.0(16)	\& 18			\\
23	&	\textbf{C$_2$P}      &	 1.0(12)    	&	1.1(09) &	8.3(09)   	&									\textbf{	-87}	&	6.3(16) \& 28   \\
24	&	\textbf{SiC}        	&	 6.0(13)    	&	1.5(13)	&	1.0(13)	&										\textbf{	50}	&	7.1(16)	\& 32	\\		
25	&	SiC$_2$    	&	 2.0(14)    	&	2.6(15) 	& 2.6(15) 	&										0		&	4.0(16)	\& 18	 & 3.6(16) \& 16		\\
26	&	SiC$_3$    	&	 4.0(12)    	&	1.7(12)	&	1.5(12)	&											13	&	4.5(16)	\& 20			\\
27	&	SiC$_4$    	&	 7.0(12)    	&	9.6(10)	&	9.5(10) 	&										1	&	5.6(16)	\& 25			\\
28	&	H$_2$CS    	&	 1.0(13)    	&	4.7(11) 	&	3.8(11)	&										24	&	5.6(16)	\& 25			\\
29	&	C$_2$S      	&	 9-15(13) 	&	1.7(13) 	&	1.4(13)	&										21	&	6.3(16)	\& 28			\\
30	&	C$_3$S     	&	 6-11(13) 	&	1.6(13)	&	1.3(13) 	&											23	&	4.0(16)		\& 18		\\
31	&	C$_3$O     	&	 1.0(12)    	&	6.4(11)	&	6.0(11)  	&										7	&	4.5(16)	\& 20			\\
32	&	H$_2$CO    	&	 5.0(12)    	&	1.9(11)	&	1.6(11)	&											19	&	5.6(16)		\& 25		\\
33	&	HCO$^+$    	&	 3-4(12)  	&	1.4(12)  	&	1.3(12) 	&										8	&	5.0(16)	\& 22			\\
34	&	C$_2$H             	&	3-5(15)	&	1.1(16)  	&	9.7(15)	&										13	&	3.5(16)	\& 16  & 3.4(16) \& 15 			\\
35	&	C$_3$H             	&	3-7(13)	&	1.5(14)	&	1.4(14)	&											7	&	3.5(16)		\& 16  & 3.3(16) \&  15		\\
36	&	\emph{c}-C$_3$H$_2$	&	2.0(13)  	&	6.1(13)	&	5.5(13)	&										11	&	3.5(16)	\& 16			\\
37	&	\emph{l}-C$_3$H$_2$ 	&	 3.0(12)    	&	1.3(13)  	&	1.1(13) 	&						18	&	2.5(16)	\& 11			\\
38	&	CH$_3$CCH          	&	1.6(13)	&	1.4(12)	&	1.3(12) 	&										8	&	2.5(16)	\& 11			\\
39	&	C$_4$H             	&	2-9(15)	&	7.2(14)	&	6.5(14)	&											11	&	2.8(16)	\& 12  & 3.4(16) \& 15 			\\
40	&	C$_5$H     	&	 2-50(13) 	&	4.5(13) 	&	4.1(13)   	&										10	&	3.5(16)	\& 16			\\
41	&	C$_6$H     	&	 7.0(13)    	&	7.3(14)	&	5.7(14) 	&										28	&	4.5(16)	\& 20	 & 3.4(16) \& 15		\\
42	&	C$_7$H     	&	 1-2(12)  	&	8.2(13)	&	7.3(13)	&												12	&	5.6(16)	\& 25			\\
43	&	\textbf{C$_8$H }     	&	 5.0(12)    	&	1.6(14) 	&	1.2(14) 	&								\textbf{33}	&	4.0(16)	\& 18			\\
44	&	C$_4$H${^-}$	&	 7.0(11)    	&	1.3(13)	&	1.3(13) 	&									0	&	2.5(16)	\& 11			\\
45	&	\textbf{C$_6$H$^-$} 	&	 4.0(12)    	&	9.8(13)	&	8.2(13) 	&										20	&	\textbf{1.3(17)}	\&\textbf{ 58}			\\
46	&	C$_8$H$^-$ 	&	 2.0(12)    	&	3.2(12)	&	2.9(12)	&											10	&	4.0(16)		\& 18		\\

\hline
\end{tabular}
} 
\tablefoot{
\tablefoottext{a} {a(b) = $a\times10^{~b}$. \textcolor{black}{Boldface values indicate column densities showing big percentage changes ($> 30\%$) between different models, in addition to the largest and smallest radii of peak abundances}}.
\tablefoottext{b} {\emph{c}-C$_3$H$_2$ and \emph{l}-C$_3$H$_2$ represent cyclic- and linear- C$_3$H$_2$}.
\tablefoottext{c} {The compiled observational results were taken from Table 7 of \cite{McElroy13}.}
\tablefoottext{d} {The most complete model (SS model with full shielding) in this work.}
\tablefoottext{e} {Model similar to that of \cite{McElroy13} with only a few modifications, e.g.,the photodissociation rates of \ce{CN-} and \ce{C3N-} have been updated. These have only   a small effect on the calculated column densities of species apart from \ce{CH3CCH} (old vs. new, 4.9 $\times 10 ^{11}$ vs. 1.4$\times 10^{12}$ cm$^{-2}$) and \ce{CH2CHCN} (3.6$\times 10^{10}$ vs. 5.3$\times 10^{11}$ cm$^{-2}$). }
\tablefoottext{f} {Comparison of computational results between the SS model and the model of \citet{McElroy13}, estimated by $100 \times (N_{\rm{This} }$ $-$ $N_{\rm{R{\tiny ATE}12}})$ $/$ $N_{\rm{R{\tiny ATE}12}}$. }
\tablefoottext{g}{The currently predicted radii of peak abundances for the species.  }
\tablefoottext{h}{Observed radii of peak emission for a few species, estimated from their interferometric maps. Observational references are given under Fig. \ref{figpeakradius}, with priority given to the newer detections. }
}
\end{table*}
}


\abstract 
{  The envelopes of AGB stars are irradiated \textcolor{black}{externally} by ultraviolet photons; hence, the chemistry is  sensitive to the photodissociation of \ce{N2} and CO, which are major reservoirs of nitrogen and carbon, respectively.
The photodissociation of \ce{N2} has recently been quantified by laboratory and theoretical studies.
Improvements have also been made for CO photodissociation. 
}
{
For the first time, we use accurate \ce{N2} and CO photodissociation rates and shielding functions in a model of the circumstellar envelope of the carbon-rich AGB star, IRC +10216.
 }  
{
  We use a state-of-the-art chemical model of an AGB envelope, the latest CO and \ce{N2} photodissociation data, and a new method for implementing molecular shielding functions in full spherical geometry with isotropic incident radiation. 
We compare computed column densities and radial distributions of molecules with observations.
}
{The transition of \ce{N2} $\to$ N (also, CO $\to$ C $\to$ \ce{C+}) is shifted towards the outer envelope relative to previous models.
  This leads to different column densities and radial distributions of N-bearing species, especially those species whose formation/destruction processes largely depend on  \textcolor{black}{the availability of} atomic or molecular nitrogen, for example, C$_n$N ($n$=1, 3, 5), C$_n$N$^-$ ($n$=1, 3, 5), HC$_n$N ($n$=1, 3, 5, 7, 9), H$_2$CN and CH$_2$CN.
 }
{The chemistry of many species is directly or indirectly affected by the photodissociation of \ce{N2} and CO, especially in the outer shell of AGB stars where photodissociation is  \textcolor{black}{important}. Thus, it is important to include \ce{N2} and CO shielding in astrochemical models of AGB envelopes and other irradiated environments.
In general, while differences remain between our model of IRC +10216 and the observed molecular column densities, better agreement is found between the calculated and observed radii of peak abundance.
 }

\maketitle
%



\section{Introduction}   \label{sec:Introduction}
\placefigurecsestructure \label{placefigurecsestructure} 

The asymptotic giant branch (AGB) is the last nuclear-burning phase for low- to intermediate-mass  stars (from 0.8 M$_\odot$ to $\sim$ 8 M$_\odot$), where the core of the star evolves into an inert and degenerate C-O core \citep{Herwig05}. AGB stars can be classified  by the elemental C/O ratio, namely, C-rich AGB stars (C/O $>$ 1), M-type AGB stars (C/O $<$ 1) and S-type AGB stars (C/O $\approx$ 1). Our own Sun will become an \textcolor{black}{M-type} AGB star.
\textcolor{black}{Generally,  \textcolor{black}{the envelopes of} C-rich stars contain CO plus C-bearing molecules, such as C$_2$, CN, HCN, and C$_2$H$_2$, while O-rich stars contain CO plus O-bearing molecules such as H$_2$O, TiO, and VO \citep{LeBertre97}.} AGB stars undergo considerable mass loss and eject dust and molecules into the surrounding regions, creating circumstellar envelopes (CSEs). These gas and dust envelopes eventually merge with the interstellar medium (ISM), enriching  molecular clouds  \textcolor{black}{in which new stars may be born}.  Furthermore, CSEs are one of the richest sources for detecting new molecules. The study of AGB stars is of particular interest and importance for our understanding of molecular formation, destruction, and the recycling of material between star birth and star death. The schematic structure of the CSE of an AGB star is shown in Fig.~\ref{figcsesturcture}.

\textcolor{black}{IRC +10216 (CW Leonis), the brightest object in the sky at mid-infrared wavelengths outside the solar system, is the nearest C-rich AGB star} and attracts intensive theoretical and observational study \citep[e.g.,][]{Morris75, Bieging88,Glassgold96,Mauron99,Millar00,Cernicharo00,Woods03,Cordiner09,De_Beck12,Agundez12,McElroy13}. More importantly, it is one of the richest molecular sources in the sky. To date, \textcolor{black}{around 180 molecular species, not counting isotopologues, have been identified in the interstellar medium or CSEs\footnote{\tt http://www.astro.uni-koeln.de/cdms/molecules/}
while more than 80 of them have been detected in IRC +10216.} Among these, more than 30 nitrogen-bearing species  \textcolor{black}{have been} identified, including \ce{HCN}, CN, SiN, PN, HNC, MgCN, NaCN, NH$_3$, HC$_7$N, and C$_5$N$^-$, amongst others \citep{Wakelam10}. Even water, which was thought unlikely to form in C-rich AGB envelopes, has been detected here \citep{Melnick01,Decin10}. Some molecular detections in IRC +10216 were their first discoveries in astrophysical environments, for example, the first detection of the cyanide anion CN$^-$ \citep{Agundez10}, and  FeCN \citep{Zack11}.  

Most of the detected species in IRC +10216 are  \textcolor{black}{found} in the cool, expanding outer CSE, where photodissociation processes dominate  \textcolor{black}{the destruction of molecules}.  \textcolor{black}{The investigation of} `key' reactions, which might affect the entire chemical network and therefore the comparison of simulations and observations,  is an important topic. 
According to a systematic sensitivity study using a large chemical network, \textcolor{black}{the reaction}     
\begin{align}
     \ce{N2 + h\nu \to N + N }     \label{eq:N2PD}
\end{align}
is one of the most significant reactions in the outer CSE, which directly affects the abundances of many N-bearing species, such as \ce{N}, \ce{N2}, \ce{HC2N}, \ce{C3N}, \ce{C3N-} and \ce{C2N} \citep{Wakelam10}. Eq. (\ref{eq:N2PD}) is the primary destruction route of N$_2$ in any region where UV photons are present. From an observational point of view, the direct detection of N$_2$ is very challenging because it has no permanent electric dipole and thus possesses no \textcolor{black}{electric-dipole-allowed} pure rotational spectrum. The only reported detection of molecular nitrogen is via its far-UV electronic transitions observed in the interstellar medium \citep{Knauth04}. One can \textcolor{black}{infer} \ce{N2} indirectly through the protonated ion, \ce{N2H+} \citep{Turner74,Herbst77} or its deuterated form, \ce{N2D+}. However, neither \ce{N2H+} nor \ce{N2D+} have been identified in IRC +10216. From the simulation point of view, Eq. (\ref{eq:N2PD}) has usually not been treated properly in models because these have not included \ce{N2} self-shielding. The importance of self-shielding of molecules in CSEs was first noticed for CO some 30 years ago \citep{Morris83}. Even with an approximate treatment  of CO self-shielding (the `one-band approximation'), much better agreement was obtained between simulations and observations. 
%

\placetabIrcParameters 

We here employ the latest reported photodissociation rate and shielding functions for \ce{N2} \citep{Li13,Heays14} to investigate the effects in a chemical model of the CSE of IRC +10216. These are based on a concerted laboratory  \citep[e.g.,][]{ajello_etal1989,helm_etal1993, sprengers_etal2004b, Stark08, lewis_etal2008a,  Heays11} and theoretical \citep[e.g.,][]{spelsberg_meyer2001,  Lewis05a,  lewis_etal2008b, ndome_etal2008} effort over the last two decades. An update was also made to the photodissociation of CO using the self-shielding functions from \cite{Visser09} following a similarly large experimental effort over the past decades. While the absolute unshielded rates \textcolor{black}{of both \ce{N2} and CO} are changed only at the level of $\sim$ 30\% \textcolor{black}{with respect to the previous values \citep{vanDishoeck88}}, it is important to realize that the uncertainty in the photorates is reduced  \textcolor{black}{by} an order of magnitude. Also, we  \textcolor{black}{develop} and employ a new fully spherically-symmetric (SS) model that computes the self-shielding of molecules in an isotropic interstellar radiation field, rather than the usual plane-parallel (PP) geometry. 

The paper is organised as follows: the CSE model, the improvements in \ce{N2} and CO photodissociation, as well as the theory and details for the evaluation of photodissociation rates for the SS model are described in Sect. \ref{sec:Methods}. The results and discussion can be found in Sect. \ref{sec:Results}, followed by the concluding remarks in Sect. \ref{sec:Concluding remarks}. The impact of these improvements on the molecular abundances in the CSE of IRC +10216 are discussed. Special attention is given to those species which have already been detected and those  \textcolor{black}{which may be} detectable in the near future, e.g., using the Atacama Large Millimeter/submillimeter Array (ALMA).  \textcolor{black}{A full description of} the SS model used for calculating the photodissociation rates of CO and \ce{N2}, together with the proposed numerical methods for implementing molecular shielding functions, are included in Appendices \ref{sec:appendix_3D_model} and \ref{sec:appendix_loop}. 
\section{Methods}                                            \label{sec:Methods}  
\subsection{CSE Model}                                       \label{sec:Model}
\placefigureT         \label{placefigureT}
\placetabInitialAB
The CSE model described in \citet{McElroy13} was employed and extended in the present work. Specific details can be found in \citet{Millar00} and  \citet{Cordiner09}. Improvements in the model are summarised in Sect. \ref{subsection:What's new}. The latest (fifth) release of the UMIST Database for Astrochemistry (UDfA), \citet{McElroy13}, hereafter R{\tiny ATE}12, was adopted in all of the calculations. R{\tiny ATE}12 contains 6173 gas-phase reactions involving 467 species. \citet{McElroy13} tested this network in a dark cloud model and  \textcolor{black}{in} a CSE model of IRC +10216 and compared the results with previous models and observations.  

The main assumptions and key parameters adopted in our model of IRC +10216 are summarised in Table \ref{tab:EnvelopeParameters}. Parent species that are injected at the inner radius of the envelope are listed in Table \ref{tab:InitialABofParentSpecies}. The simulations start at the inner envelope with the following set of parameters:  \textcolor{black}{the} radius, molecular hydrogen number density, visual extinction, and kinetic temperature of the gas  \textcolor{black}{are} $r_1$= 1.0 $\times$ 10$^{15}$ cm, $n({\rm H_2})$ = 1.3 $\times$ 10${^7}$cm$^{-3}$,  $A_{\rm V}$  = 13.8 mag, and $T$ = 575\,K, respectively. The outer radius of the envelope is set to $r_f$ = 7 $\times$ 10$^{17}$ cm, where the density has decreased to $n({\rm H_2})$ = 26 cm$^{-3}$,  \textcolor{black}{the temperature to} $T$ = 10\,K, and  \textcolor{black}{the visual extinction to} $A_{\rm V}$  = 0.02 mag. We pay particular attention to the chemistry in the outer CSE where photodissociation is  \textcolor{black}{an important process}. For all photoreactions, species are destroyed by photons from all directions, i.e., the radiation field is isotropic. For IRC +10216, one has to only take into account photons that come from the interstellar medium. The star, with a temperature of $\sim$ 2330 K \citep{De_Beck12}, is too cool to generate photons that could lead to the photodissociation of molecules. 

\subsection{What's new}    \label{subsection:What's new}
Our aim is to improve the treatment of N$_2$ and CO photodissociation. In the work of \citet{McElroy13}, they employed the following parameters and methods: 
\begin{enumerate}[(a)] 
     \item Unshielded photodissociation rates   \\
           CO: $2.0 \times 10^{-10}$ s$^{-1}$, from \citet{vanDishoeck88}\\
           N$_2$: $2.0\times 10^{-10}$ s$^{-1}$, from \citet{vanDishoeck88}   
     \item Shielding functions\\
           CO: dust + self-shielding, from \citet{Morris83}, neither taking into account all lines for CO self-shielding nor shielding from H$_2$\\
	       N$_2$: dust shielding only                                                        
	 \item Method for implementing self-shielding functions \\
	       The `one-band approximation' \citep{Morris83} 
\end{enumerate}
In this work, we employ the most accurate molecular data to date and develop a new method for implementing molecular shielding functions in the SS model:
\begin{enumerate}[(a)] 
     \item Unshielded photodissociation rates \\
     CO: $2.6 \times 10^{-10}$ s$^{-1}$, 30\% higher, from \citet{Visser09} \\
     N$_2$: $1.65\times 10^{-10}$ s$^{-1}$, 28\% lower, from \citet{Li13}                  
     \item Shielding functions \\
      CO: dust + self- + H + H$_2$ shielding, from \citet{Visser09} \\
      N$_2$: dust + self- + H + H$_2$ shielding, from \citet{Li13}     
	 \item Method for implementing self-shielding functions\\
      By iterating the calculations to get converged results for the column densities in spherical symmetry for an isotropic radiation field. Details of this method are described in Sect. \ref{subsection:pd rate and Theta} and Appendix~\ref{sec:appendix_loop}.  
\end{enumerate}
\subsection{Photodissociation rate and shielding function} \label{subsection:pd rate and Theta}
The definitions of the photodissociation rate, $k$, unshielded rate, $k^0$, and shielding function, $\Theta$, appropriate for a plane-parallel model (only considering photons from the normal direction) are the same as those described in our recent paper \citep{Li13}. In this work, we consider photons from all directions in space (SS model) for photoreactions that occur in the outer CSE. Suppose an incident ray, inclined with angle $\vartheta$ to the outward normal direction, arrives at a specific radius, $r_i$. The total column density of H$_2$ integrated from this point along the ray to infinity can be written as \citep{Jura81,Morris83}
\begin{align} \label{eq:N_H2_r_theta_Av}	
 N_{\text{H}_2}(r_i,\vartheta) = N_{\text{H}_2}(r_i, 0)~\vartheta / \sin{\vartheta}~~\text{cm}^{-2}~.
\end{align} 
\textcolor{black}{This expression diverges for $ \vartheta $ close to zero and $\pi$. In practice, the value of $ \vartheta /\ \rm{sin}\vartheta $ equals 1 if $ \vartheta = 0 $. For the case of $ \vartheta = \pi $, its value is obtained using a linear extrapolation from the preceding data points on our angular grid. }

In the CSE of IRC +10216, most hydrogen is locked in H$_2$ due to self-shielding. Thus, the corresponding visual extinction, $A_{\rm V}(r_i,\vartheta)$, is  
\begin{align} \label{eq:Av_r_theta}	
 A_{\rm V}(r_i,\vartheta) = 2N_{\text{H}_2}(r_i,\vartheta)/(1.87 \times 10^{21})~~\text{mag} ~,
\end{align}
with the conversion factor based on the observations of \citet{Bohlin78} and \citet{Rachford09}. The photodissociation rates of most interstellar species are mainly attenuated by dust, which is characterised by dust shielding functions given by
\begin{align} \label{eq:Theta_dust}	
   \Theta_\text{dust}(r_i, \vartheta) = e^{- \gamma  A_{\rm V} (r_i,\vartheta) }~,
\end{align}
where $\gamma$ is the parameter used to estimate the increased dust extinction at ultraviolet wavelengths. In general, the contribution of a specific ray of interstellar photons to the overall photodissociation rate at $r_i$ may be calculated using   
\begin{align} \label{eq:k_r_theta_dust}	
   k(r_i, \vartheta) = k^0  ~ \Theta_\text{dust}(r_i,\vartheta)~~ \text{s}^{-1}~.
\end{align}
However, the photodissociation rates of N$_2$, CO, and H$_2$ are more complex due to self- and mutual-shielding by H, \ce{H2}, CO, and other molecules (which are wavelength and column density dependent) and also continuum shielding by dust. A convenient way to take these effects into account in models is by use of molecular shielding functions that are calculated from high accuracy wavelength-dependent absorption cross sections. There is no need to consider H$_2$ photodissociation in the current study because it is fully shielded everywhere in the CSE. For N$_2$ and CO, one must consider self-shielding and mutual-shielding by \ce{H2}. The photodissociation rate of N$_2$ is calculated by    
\begin{align} \label{eq:k_r_theta_mol}	
  \textcolor{black}{  k_{\ce{N2}} } (r_i, \vartheta) =\textcolor{black}{  k^0_{\ce{N2}} } \Theta_{\rm dust}(r_i,\vartheta)~ \Theta_\text{mol}[N_{\text{H}_2}(r_i,\vartheta),N_{\text{N}_2}(r_i,\vartheta)]~~ \text{s}^{-1},
\end{align} 
\textcolor{black}{where} $k^0_{\ce{N2}}$ is the unshielded photodissociation rate of \ce{N2}, and $ N_{\ce{N2}} (r_i,\vartheta)$ is the column density of \ce{N2} at radius $r_i$, integrated from this point along angle $\vartheta$ to infinity. $\Theta_\text{mol}[N_{\text{H}_2}(r_i,\vartheta),N_{\text{N}_2}(r_i,\vartheta)]$ is the molecular shielding function that is responsible for the multiple shielding effects caused by N$_2$ (self-) and H$_2$. It can be obtained via a bi-linear interpolation of N$_2$ and H$_2$ shielding functions that depend only on their column densities. 

\textcolor{black}{Correspondingly, the CO photodissociation rate is calculated using }
\begin{align} \label{eq:k_r_theta_mol_CO}	
  \textcolor{black}{ k_{\rm CO}(r_i, \vartheta) = k^0_{\rm CO}  \Theta_{\rm dust}(r_i,\vartheta) \Theta_\text{mol}[N_{\text{H}_2}(r_i,\vartheta),N_{\text{CO}}(r_i,\vartheta)]~ \text{s}^{-1},}
\end{align} 
\textcolor{black}{where} $\textcolor{black}{k^0_{\rm CO}}$ \textcolor{black}{is the CO unshielded photodissociation rate, and} $\textcolor{black}{ N_{\rm CO}(r_i,\vartheta)} $  \textcolor{black}{is the column density of CO at radius } $\textcolor{black}{r_i}$\textcolor{black}{, integrated along angle}  $\textcolor{black}{\vartheta} $ \textcolor{black}{to infinity.} \textcolor{black}{Atomic hydrogen can also shield \ce{N2} and CO, although in practice this is unimportant for the case of CSEs because the majority of hydrogen is locked in \ce{H2}.}

\textcolor{black}{A full description of} the SS model and the numerical method for the computation of $N(r_i,\vartheta)$ can be found in Appendix \ref{sec:appendix_3D_model}. The implementation of N$_2$ and CO shielding functions in the SS model is not straightforward. This is because before starting the evolution of the chemistry, one needs to know \textcolor{black}{in advance} the abundances of N$_2$ and CO at each radius to evaluate their column densities, and to compute the molecular shielding functions. However, these data are generated from the output of the model. One approximate solution is to assume that the N$_2$ and CO number densities are \textcolor{black}{a constant proportion of} H$_2$, which falls as $r^{-2}$ \citep{Jura81}, then work out the corresponding column densities. This assumption is inaccurate in the photon-dominated regions where N$_2$ and CO densities are no longer constant with respect to \ce{H2}. In this study, we propose a new procedure \textcolor{black}{which gives} more accurate and convergent results by iterating the computation of the chemistry. This method is described in Appendix \ref{sec:appendix_loop}. 

The photodissociation of N$_2$ and CO becomes considerable at the edge of the CSE, where \textcolor{black}{the visual extinction is less than 1.0 mag}  and the temperature is around 10\,K, as can be seen in Fig. \ref{figT}.
When calculating the N$_2$ shielding functions, the excitation temperature for both H$_2$ and \ce{N2} are chosen to be 10\,K. 
The Doppler widths of N$_2$, H$_2$, and H are 0.2, 3, and 5 km s$^{-1}$, respectively. The column density of atomic hydrogen is taken to be $1.0 \times 10^{14}$ cm$^{-2}$. For CO shielding, the excitation temperature for H$_2$ and CO are chosen to be 11 and 5~K, respectively. \textcolor{black}{At the low densities in the outer CSE, CO is subthermally excited.} The Doppler widths of CO, H$_2$, and H are 0.3, 3, and 5 km s$^{-1}$. Electronic tables of the \textcolor{black}{\ce{N2} and CO shielding functions are available online.}\footnote{\tt http://home.strw.leidenuniv.nl/$\sim$ewine/photo/} 
\textcolor{black}{Note that these shielding functions were calculated by considering (\ce{H2}, H, and self-) shielding at various temperatures for \textcolor{black}{a few typical column densities}. Shielding functions for other \textcolor{black}{temperatures and column densities} can be obtained by interpolation. } In practice, the uncertainties in the shielding functions \textcolor{black}{over a temperature variation of a few degrees are negligible,} see \citet{Li13}. \textcolor{black}{The parameter $\gamma$ used to account for the dust extinction at ultraviolet wavelengths for \ce{N2} and CO was taken to be 3.9 and 3.5, respectively.}

Finally, the radial photodissociation rate \textcolor{black}{in a spherical geometry with isotropic incident radiation} is evaluated by
\begin{align} \label{eq:k_effective}	
k(r_i) = \dfrac{1}{2} \int_0^{\pi} \! k(r_i,\vartheta) \sin{\vartheta} \, \mathrm{d}\vartheta ~~ \text{s}^{-1}.
\end{align} 
\section{Results}                \label{sec:Results} 
\subsection{Improvements in N$_2$ and CO photodissociation}  \label{subsec:Improvements in N$_2$ and CO
photodissociation} 
\placefigureaDvscD         \label{placefigureaDvscD}  
\placefigurediffshielding  \label{placefigurediffshielding} 
Figure~\ref{figaDvscD} compares the calculated column densities and photodissociation rates \textcolor{black}{for the} spherically-symmetric and plane-parallel models. 
For the PP case, we consider photons from the radial direction only but the full shielding (dust + self- + H + \ce{H2}) was included. 
\textcolor{black}{The envelope was illuminated by a \citet{Draine78} field, extended to
wavelengths longer than 2000 $\AA{}$ according to 
\citet{vanDishoeck88CO} and with a scaling factor, $\chi$. In the SS model, photons from all directions in space were
considered with $\chi=1$ and the angular weighting scheme is given by
Eq. (\ref{eq:k_effective}). 
For the PP model, radiation from one side
was taken into account explicitly with $\chi=0.5$ and radiation from the far side assumed completely attenuated. Hence, in the absence of the envelope the
same unattenuated \citet{Draine78} radiation field is recovered as for
the SS model. 
Indeed, for $\chi=0.5$, the \ce{N2} photodissociation rates in
the SS and PP cases are very similar (middle panel Fig.~\ref{figaDvscD}), whereas when assuming $\chi=1.0$, the PP rate is up to a factor of two higher at the outer
edge.
As a result, the PP model with $\chi=1.0$ underestimates the \ce{N2} column density at this location, but the $\chi=0.5$ model results
are very close to those of the SS model.  The bottom panel of Fig. 3
shows that the PP and SS models start to deviate more in the
inner envelope, where the PP model gives a much higher
photodissociation rate. However, the absolute photodissociation rates
are so small here that they are negligible in the chemistry. Overall,
we conclude that for the CSE chemistry of IRC +10216, the PP model with
a scaling factor of $\chi=0.5$ provides excellent agreement with the full SS
model.}


\textcolor{black}{In the SS model the abundance of \ce{N2} is higher than that in the PP model \textcolor{black}{in the outer envelope}.  The distribution of CO in the SS and PP models is similar to that for N$_2$, also indicating a higher abundance of CO at the edge of the cloud for the SS case.}

Including or excluding different shielding effects in the model is another crucial factor in the investigation of N$_2$ and CO photodissociation. Shielding effects, which act like a `smoke screen', help molecules survive in photon dominated regions. N$_2$ and CO can be significantly shielded by both dust (wavelength independent) and molecules (wavelength dependent, line-by-line shielding). Fig.~\ref{figdiffshielding} presents the various shielding effects on N$_2$ photodissociation. Self-shielding is the most significant shielding effect, whereas (H$_2$ + H) shielding contributes the least. Shielding from dust largely depends on the properties of the dust \citep{vanDishoeck06}. It is necessary to include \textcolor{black}{all sources of shielding for \ce{N2} because the combined effect has a strong influence on the abundance and distribution of \ce{N2}.}  
\placefigureNb     \label{placefigureNb} 
\placefigureCO     \label{placefigureCO} 

Figure~\ref{figNb} shows the fractional abundances of N$_2$ and N, calculated using various models. The location of the transition zone from N$_2$ to N is \textcolor{black}{shifted outwards by a factor of 5} when molecular shielding is taken into account in the PP ($\chi=1.0$) model. Further changes happen when we use the SS model and PP ($\chi=0.5$) model.

The photodissociation of CO has an impact on both carbon and nitrogen chemistry. In particular, reactions with C$^+$ are important destruction mechanisms for quite a few N-bearing species, for example C$_n$N$^{-}$ ($n$=1, 3, 5). As shown in Fig.~\ref{figCO}, the conversion of CO to C and \ce{C+} is affected when \textcolor{black}{CO self-shielding is calculated using the updated shielding functions}. The chemistry of other C-containing molecules is also affected by the photodissociation of CO, but are beyond the scope of this work. The SS model with full shielding predicts that CO is relatively abundant up to a radius of $\sim6\times10^{17}$ cm, which is in better agreement with observations than the PP model (which is described further in Sect. \ref{subsubsec:Comparison with observations:radius of peak abundance}). 
\subsection{Impacts of N$_2$ and CO photodissociation on the CSE chemistry}    \label{subsec:impacts of N$_2$ and CO PD positions} 
The photodissociation of N$_2$ and CO significantly affects the chemistry in the outer CSE. This influence has two observable aspects: one leads to changes in the radial distribution of  the species' peak abundances. 
These distributions will become increasingly observable \textcolor{black}{after exploiting the increased spatial resolution and sensitivity of new facilities such as ALMA.}
The other effect leads to changes in the peak abundances, which are directly reflected in total column densities. 
In the following subsections, we discuss fractional abundances of the most interesting (detectable or promisingly detectable) species which are sensitive to the photodissociation of N$_2$ and CO. Special attention is given to N-bearing species. 
Unless stated elsewhere, comparisons are made between the SS model including the updated \ce{N2} and CO shielding and a recreation of the model of \citet{McElroy13}.
Additional comparisons between observations and simulations from this study and \citet{McElroy13} \textcolor{black}{ with regards to the 46 detected species in IRC +10216 are discussed in  Sect. \ref{subsec:Overview of the changes in the total column densities of the detectable species}.}
   
The dominant formation and destruction mechanisms of most species vary with physical conditions in the CSE. Here, we mainly focus on the chemistry of the outer CSE where photodissociation processes dominate. 

\subsubsection{CN$^-$, C$_3$N$^-$ $\&$ C$_5$N$^-$}
 \label{subsubsec:CnN-} 
\placefigureCaceNy    \label{placefigureCaceNy} 
When N$_2$ photodissociation is treated in an appropriate manner, \textcolor{black}{a higher abundance of} N$_2$ is present in the outer CSE with a corresponding decrease in N. The most direct effects are on those N-bearing species whose formation processes require atomic N. \ce{C_nN-} ($n$=1, 3, 5) have already been observed in the CSE of IRC +10216 \citep{Agundez10,Thaddeus08,Cernicharo08}. The following discussion of \ce{C_nN-} chemistry is based on the adopted reaction network from \cite{Walsh09} which was incorporated into \citet{McElroy13} and our model. Many of the adopted rate coefficients come from \citet{Eichelberger07}.     

The chemistry of the molecular anions, \ce{CN-} and \ce{C3N-}, in the CSE of IRC +10216 has been described in \citet{Kumar13}. As shown in Fig~\ref{figCaceNy}, C$_n$N$^-$ ($n$=1, 3, 5) are very sensitive to \ce{N2} photodissociation and are significantly decreased in the new treatment. This is because their formation processes are strongly affected by the abundance of atomic N in the outer CSE. As an example, at 1.0 $\times$ 10${^{17}}$ cm \ce{CN-} and \ce{C3N-} are mainly formed through:
\begin{align} \label{eq:CN-formation}
& \text{N} + \text{C}{_n^-} \to \text{CN}{^-} + \text{C}{_{n-1}} ~~(n= 5, 6, 7, 8, 9, 10),  
\end{align}
\begin{align}\label{eq:C3N-formation}
& \text{N} + \text{C}{_n^-} \to \text{C}{_3}\text{N}{^-} + \text{C}{_{n-3}} ~~(n= 6, 7, 8, 9, 10).
\end{align}
The abundances of these anions decrease with the decreasing abundance of N. As shown in Fig. \ref{figNb}, the fractional abundance of N drops from $4\times10^{-4}$ (\cite{McElroy13} model) to $3\times10^{-5}$ (SS model) at 10$^{17}$ cm, leading to $\sim$ 1 order of magnitude drop \textcolor{black}{in} \ce{CN-} and \ce{C3N-}.
Interestingly, C$_5$N$^-$ has a different formation route. It comes primarily from C$_5$N, via
\begin{align} \label{eq:C5N-formation}
& \text{C}{_5}\text{N} + e{^-} \to \text{C}{_5}\text{N}{^-}  + h\nu.  
\end{align}
The radiative electron attachment of \ce{C5N} in Eq. (\ref{eq:C5N-formation}) is very efficient compared with \ce{C3N}, which leads to the higher abundance of \ce{C5N-} \citep[see, e.g.,][]{Herbst08}. 


\subsubsection{CN, C$_3$N $\&$ C$_5$N}
\placefigureCaceN     \label{placefigureCaceN}
The chemistry of C$_n$N ($n$=1, 3, 5) are also sensitive to N$_2$ photodissociation, but not as sensitive as C$_n$N${^-}$ ($n$=1, 3, 5). This is because they are only partly affected by  the available abundance of atomic N. Take CN as an example, it is mostly formed by the following reactions,
\begin{align} 
&\text{HCN} + h\nu    \to \text{CN} + \text{H},  \label{eq:HCN+hv} \\
&\text{N} + \text{C}{_2}    \to \text{CN}  + \text{C}, \label{eq:CNformation} \\  
&\text{N} + \text{CH} \to \text{CN} + \text{H}.  
\end{align}
The photodissociation of HCN dominates CN formation inside a radius of $\sim$ 4$\times10^{16}$ cm. Further out, the reactions involving atomic N play a role, although these reactions do not affect its peak abundance, see Fig.~\ref{figCaceN}. The cases of C$_3$N and C$_5$N are similar to CN.   
 
There is a shift in the peak abundances of CN, \ce{C3N-} and C$_5$N$^-$ from the outside of the envelope inwards towards the star. In the case of \ce{C5N-}, this shift is from 10$^{17}$ to $4\times 10 ^{16}$ cm. This difference is caused by the change in location of the \ce{N2} to N transition zone when \textcolor{black}{full shielding is included}.

\subsubsection{{C$_7$N} $\&$ \text{C$_9$N}}
\placefigureCgiN      \label{placefigureCgiN} 
As shown in Fig.~\ref{figCgiN}, the peak abundances of C$_7$N and C$_9$N are located at radii between 5 $\times$ 10$^{16}$ and 1 $\times$ 10$^{17}$ cm. Formation mechanisms for C$_7$N and C$_9$N are thought to be very similar to those of C$_n$N ($n$=1, 3, 5). They have a formation route via:
\begin{align} \label{eq:C9Nformation}
&\text{N} + \text{C}{_7^-} \to \text{C}{_7}\text{N}  + e{^-}, \\   
&\text{N} + \text{C}{_9^-} \to \text{C}{_9}\text{N}  + e{^-}.   
\end{align}
These larger C$_n$N molecules have not yet been identified in IRC +10216, but they are also sensitive to N$_2$ photodissociation and may be useful in indirect investigations of the chemistry and abundance of \ce{N2}. Their predicted peak abundances are around one order of magnitude lower than that of \ce{C5N.}

\subsubsection{HC$_n$N ($n$ = 3, 5, 7, 9)}
\placefigureHCacegiN       \label{placefigureHCacegiN}  
The cyanopolyynes, \ce{HC_nN} ($n$ = 3, 5, 7, 9), are also influenced by N$_2$ photodissociation, especially in the outer shell where photodissociation processes are important, as shown in Fig.~\ref{figHCacegiN}. The fractional abundances of HC$_n$N ($n$ = 3, 5, 7, 9) are  all affected when the full shielding of \ce{N2} is considered. The most significantly affected species is HC$_5$N. The changes in their total column densities are rather small because the peak abundances, which contribute most to the column densities, are unchanged. 
However, the radial distribution of \ce{HC5N} may be spatially resolvable by sensitive interferometers. 

The chemistry of large organic molecules is more complex than for simple species. At the photon-dominated radius $\sim$~10$^{17}$ cm, the main destruction process for the parent species, HCN, is photodissociation, which gives rise to its daughter molecule CN. Another important process for the destruction of HCN is by C${^+}$, via
\begin{align} \label{eq:HCNwithC+}
& \text{HCN} + \text{C}{^+} \to \text{CNC}{^+}  + \text{H}.    
\end{align} 
However, there are quite a few reformation reactions at this location. Amongst the most important are two reactions relevant to the present study,
\begin{align} \label{eq:HCNformation}
& \text{N} + \text{CH}{_2} \to \text{HCN}  + \text{H},   \\
& \text{H} + \text{CN}{^-} \to \text{HCN}  + e{^-}. 
\end{align} 
The decrease of HCN in the outer CSE, shown in Fig. \ref{figHCacegiN}, is largely due to the enhanced shielding of N$_2$, through the decreased abundance of N and, indirectly, \ce{CN-}. 

In our calculations, HC$_n$N are synthesised by the reactions
\begin{align} \label{eq:HC379Nformation}
& \text{HC}{_{n-1}}\text{H} + \text{CN} \to \text{HC}{_n}\text{N}  + \text{H} ~~(n=3, 7, 9).
\end{align} 
The changes in the radial distributions for HC$_n$N ($n$=3, 7, 9) are very similar to those of CN, see Fig.~\ref{figCaceN}.  
HC$_5$N is an exception, and has the largest change among all HC$_n$N species in the photodissociation region. This is because the main formation process for HC$_5$N is via associative electron detachment,
\begin{align} \label{eq:HC5Nformation}
& \text{H} + \text{C}{_5}\text{N}{^-} \to \text{HC}{_5}\text{N}  + e{^-}.
\end{align} 
As discussed previously, the abundance and distribution of \ce{C5N-} is strongly affected by the inclusion of \ce{N2} shielding, thus, the abundance of \ce{HC5N} is also affected.

\subsubsection{H$_2$CN $\&$ CH$_2$CN}
\placefigureHbCN      \label{placefigureHbCN} 
The H$_2$CN and CH$_2$CN peak abundances are also significantly decreased when full N$_2$ shielding is included, as shown in Fig.\ref{figHbCN}. Their main formation processes in the outer CSE are
\begin{align} \label{eq:H2cNformation}  
& \text{N} + \text{CH}{_3} \to  \text{H}{_2}\text{CN} +  \text{H}
\end{align}
and
\begin{align} \label{eq:H2cNformation}  
& \text{N} + \text{C}{_2}\text{H}{_3} \to \text{CH}{_2}\text{CN} + \text{H}.     
\end{align}
There has already been a detection of CH$_2$CN in IRC +10216 \citep{Agundez08}, whereas H$_2$CN has only been detected in dark clouds \citep{Ohishi94}. Since the column density computed here is close to the value observed in dark clouds, this species could also be detectable in IRC +10216. Because the chemistry of the two species are relatively simple and largely depend on the availability of N, we may be able to use them to constrain the fractional abundance of N$_2$ in space by comparing observations with models.
\subsubsection{PN}
\placefigurePN  
The chemistry of P-bearing and Si-bearing species in C-rich CSEs are not completely understood \citep{McElroy13}. The most sensitive P-bearing and Si-bearing species to \ce{N2} shielding are HNSi, SiN, and PN, see Fig. \ref{figPN}. Here we concentrate only on the chemistry of PN, which is mainly formed through the following route at a radius of $10^{17}$ cm:  
\begin{align} \label{eq:PNformation}  
& \text{N} +   \text{PH}     \to      \text{PN}  +   \text{H}.
\end{align}
This process is largely reliant on atomic nitrogen, which is significantly decreased at the PN peak radius when updated shielding is used.
\subsection{Comparison with observations} 
\label{subsec:comparison with observations}
\label{subsec:Overview of the changes in the total column densities of the detectable species}
\placetabNcol      \label{placetabNcol}  
\subsubsection{Column density} \label{Comparison with observations:column density} 
The radially-integrated column densities of most species are controlled by their peak abundances. After investigation, the abundances of quite a few of the 467 species in the network are affected when we use an appropriate method to calculated the photodissociation of N$_2$ and CO. A summary of the changes in column densities is an efficient way to extract the effects of these updates on the chemistry in the CSE. Here, we concentrate on \textcolor{black}{those} 46 species which have been observed in IRC +10216. The calculated and observed column densities are listed in Table~\ref{tab:Ncol}. Of these, the calculated values for 14 species vary by more than 30\%, and the values for 6 species vary by more than 80\%, when compared with results from \cite{McElroy13}. The most affected N-bearing species with respect to the previous model are \ce{CN-} (-91\%), SiN (-85\%), and \ce{CH2CN} (-93\%). 

All of the species listed in Table~\ref{tab:Ncol} contain nitrogen and/or carbon and their formation and destruction may be influenced by \ce{N2} and CO photodissociation. The important reactions for some of these species have been discussed in the previous subsections. The most significantly altered column densities include \ce{CN-} and \ce{C3N-}, whose formation is sensitive to the available N, which is reduced by the increased shielding of \ce{N2} in our new calculations. A factor-of-10 decrease for SiN and \ce{CH2CN} column densities can be similarly explained.   

Some of these changes reduce the agreement of the CSE chemistry model with observed column densities. There are many possible reasons for the discrepancies. First of all, some of the assumptions in \textcolor{black}{the physical model of the} CSE, unrelated to photodissociation (see Table \ref{tab:EnvelopeParameters}), may be too simple. Further refinement of these assumptions may improve the agreement. For example, in this study, we ignored the presence of gas and dust shells in the \textcolor{black}{expanding envelope}. According to the study of \cite{Cordiner09}, the inclusion of these shells gives a significant improvement in both modelled column densities and spatial distributions. A study of photodetachment as a destruction mechanism for the N-bearing anions, \ce{CN-} and \ce{C3N-}, in the CSE of IRC +10216, concluded that the inclusion of shells with enhanced density, similar to the model of \cite{Cordiner09}, increases the column densities of the anions by about 20\% \citep{Kumar13}.


One of the challenges in CSE simulations is the uncertainty in the abundance of parent species (see Table \ref{tab:InitialABofParentSpecies}). 
In particular the modelled column densities of Si-bearing and P-bearing species in Table~\ref{tab:Ncol} are several orders of magnitude lower than that suggested by observations. 
Possibly, the currently modelled Si-bearing and P-bearing parent species should have a higher abundance or there exist other Si- and P-containing parent species.
On the other hand, the uncertainties of some observations are considerable. 
Take \ce{SiC2} as an example. The cited column density for this molecule in Table \ref{tab:Ncol} is 2.0 $\times10^{14}$ cm$^{-2}$  \citep{Thaddeus84}. However, according to a more recent study by \cite{Cernicharo10} employing high quality Herschel/HIFI spectra an average column density of $\sim$ 6.4$\times10^{15}~ \rm{cm}^{-2}$ was found.  \textcolor{black}{ Our computed value, $ 2.6\times10^{15}$ cm$^{-2}$, is in better agreement with the latter observed value.  }       
\placefigurepeakradius
\subsubsection{Radius of peak abundance} \label{subsubsec:Comparison with observations:radius of peak abundance} 
The radial location of  \textcolor{black}{peak molecular abundance} is another observable parameter and is  \textcolor{black}{of equal importance to} the column densities. Many daughter species have a clear radial peak in their modelled abundances which we may directly compare with spatially resolved observations. A selection of such observations have been reduced to their peak radii (without considering detailed excitation or radiative transfer) and compared with our model calculations, and are summarised in Table~\ref{tab:Ncol} and Fig.~\ref{figpeakradius}. 

The model predicts that most peak radii fall within shells between $1\times 10^{16}$ (4$^{\prime \prime }$) and $2 \times 10^{17}$ cm (89$^{\prime \prime }$) from the central star. Here, the distance between the Earth and IRC +10216 is assumed to be 150 pc, based on the study by \citet{De_Beck12}. Different values for this parameter have been employed in other work which may affect intercomparison, e.g., \cite{Huggins88} who assumed 200 pc, and \cite{Cordiner09} who employed 130 pc.

The outer radius of CO using the SS model is slightly larger, $\sim$ 7 $\times$ 10$^{17}$ cm ($\sim 310^{\prime \prime}$), than the previous value, see Fig.~\ref{figCO}. 
Emission from CO in IRC +10216 has been extensively investigated. 
A good summary can be found in \citet{De_Beck12}, who also obtained various physical parameters of the structure of IRC +10216. However, conclusions from these observations are not always the same. For instance, \citet{Huggins88} detected CO emission out to $\sim 210''$ in the $J=1-0$ line and 180$''$ in the $J=2-1$ line. Later, the $J=1-0$ line was detected out to 190$''$ \citep{Fong03, Fong06}. Recently, Cernicharo et al. (private  communication)  observed weak $J=2-1$ emission out to a radius of $\sim 300^{\prime \prime}$, in excellent agreement with the present simulations. 

Figure~\ref{figpeakradius} compares the simulated and observed peak abundance radii for a few selected species  \textcolor{black}{for which} interferometric maps are available. The observed peak radii, $R_{\rm obs}$, were deduced from direct measurement of the peak radius and averaged  \textcolor{black}{over those maps for which} multiple rotational transitions were available. These observations were done with high angular resolution telescopes, such as the Plateau de Bure Interfrometer (PdBI) \citep{Guelin98, Guelin99, Guelin11} and the Very Large Array (VLA) \citep{Dinh-V-Trung08}. In our models, it is assumed that the peak emission occurs at the peak abundance position, i.e., no excitation or radiative transfer of the model lines is performed. 

With these assumptions, it is seen in Fig.~\ref{figpeakradius} that the overall agreement between our simulations and observations is within a factor of two.
Interestingly, the simulations predict a peak abundance at 18$^{\prime \prime}$ for the parent species \ce{SiC2} (see Table \ref{tab:Ncol}). 
This is very close to the value estimated from observations \citep{Lucas95,Guelin11}, 16$^{\prime \prime}$.
No other parent species exhibits a similar maximum.

None of the peak radii of observationally-mapped species were found to be significantly altered by the introduction of a more appropriate photodissociation treatment, even \textcolor{black}{ if} their column densities were substantially changed. However, large differences were deduced for some unmapped carbon chain molecules and anions, specifically \ce{C3N-}, \ce{C5N-}, \ce{C5N}, \ce{C7N}, and \ce{C9N}. The peak radii of these species all shift inwards after applying the new photodissociation shielding functions. For the case of \ce{C5N-}, the peak radius retreats by a factor of 2.5, from $1.0 \times10^{17}$ (44$^{\prime \prime}$) to $4.5 \times10^{16}$ (20$^{\prime \prime}$) \,cm. 

Our new model predicts that the \ce{CN-} peaks at a somewhat larger radius than its corresponding neutral CN, whereas the peaks of \ce{C3N-} and \ce{C5N-} match those of \ce{C3N} and \ce{C5N}, respectively, as shown in Table~\ref{tab:Ncol}, Fig. \ref{figCaceNy}, and Fig. \ref{figCaceN}. This conclusion differs from some previous models, e.g., \citet{Kumar13} and references therein. Similar differences were found between the modelled peak radii of the neutral and anionic forms of \ce{C_nH} (n=2, 4, 6) (Table~\ref{tab:Ncol}). Interferometric maps of these species could  serve as a test of the model predictions. 



\subsection{N$_2$ abundance in IRC +10216}               \label{subsec:N$_2$ abundance} 
\placefigurediffNb

There is some uncertainty in the initial abundance of N$_2$ (a parent species) in the CSE of IRC +10216. In previous models, its adopted abundance relative to \ce{H2} \textcolor{black}{lay} between 1 $\times 10^{-5}$ \textcolor{black}{and} $2 \times 10^{-4}$ \citep[e.g.,][]{Nejad87, Millar00, Mackay01,Agundez06}. 
In the present work, we employ the same abundance as that in the model of \cite{McElroy13}, 2 $\times$ 10$^{-4}$, which gives an overall good agreement between the modelled and observed N-bearing species' abundances in Table \ref{tab:Ncol}. 
However, the observational investigation of  \cite{Milam08} deduced an N$_2$ abundance of $\sim1\times10^{-7}$. 
They used the observation of PN to infer an \ce{N2} abundance by assuming approximate equality of the abundance ratios, PN/\ce{N2} $\simeq$ [P]/[N] and [P]/[N] $\simeq$ HCP/HCN $\simeq$ 0.001. 

To test the sensitivity of daughter species to the initial \ce{N2} abundance we recomputed their chemical abundances after  \textcolor{black}{lowering} the parent \ce{N2} abundance to $1\times 10^{-5}$, a factor of 20 lower than in our  \textcolor{black}{fiducial} models.
The results for four species, \ce{CN-}, \ce{C5N-}, \ce{C5N-}, and \ce{N2H+} are shown Fig. \ref{figdiffNb}.

\ce{N2H+} has not been observed in IRC +10216, but has been detected in other interstellar and circumstellar environments \citep[e.g.,][]{Liu13,Qi13,Codella13}.
This species is thought to be a good tracer for N$_2$, from which it forms directly via the reaction,
\begin{align} \label{eq:N2Hxformation}  
& \text{N}{_2} +   \text{H}{_3^+}   \to  \text{N}{_2}\text{H}^{+}  +  \text{H}.
\end{align}
The utility of \ce{N2H+} as an \ce{N2} tracer is supported by the calculated abundances plotted in Fig.~\ref{figdiffNb}, which are directly proportional to the assumed \ce{N2} abundance.
The predicted column density of \ce{N2H+} assuming initial \ce{N2} abundances of $2\times10^{-4}$ and $1\times10^{-5}$ are 9.4 $\times$ 10$^{10}$ and 5.3 $\times$ 10$^{9}$ cm$^{-2}$, respectively. The peak abundance of \ce{N2H+} is located at a radius of $5\times10^{16}$ cm (22$^{\prime \prime }$).   
Future deeper observations of IRC+10216 with ALMA may determine the column density of \ce{N2H+} and allow for a new estimate of the \ce{N2} abundance. Note that the abundances of neither \ce{N2H+} nor \ce{HCO+}, formed through a similar reaction between CO and \ce{H3+}, are \textcolor{black}{significantly} affected by the new model treatment. 

The abundances of \ce{C_nN-} anions are \textcolor{black}{also} not significantly affected by the altered parent \ce{N2} abundance. 
These molecules are strongly dependent on the availability of atomic nitrogen (Sec.~\ref{subsubsec:CnN-}) which in the region of their peak abundance forms primarily from the photodissociation of HCN.
Thus, HCN is the important N-bearing parent species affecting their column densities. \textcolor{black}{Note that the main source of atomic N further out in the envelope is via the photodissociation of \ce{N2}.  }

\section{Concluding remarks}   \label{sec:Concluding remarks}

In this work, the effect of self-shielding of \ce{N2} and CO, and mutual-shielding by \ce{H2}, in the outer envelope of the C-rich AGB star, IRC +10216, was studied for the first time.
This was performed using the latest available data for \ce{N2} and CO photodissociation, and using an extended spherically-symmetric model for molecular self-shielding. 
The impact of these two improvements on the chemical evolution of the expanding envelope was investigated, with special attention paid to nitrogen chemistry and detectable species. 

Key points from this study are:
\begin{enumerate}[(i)] 
\item N$_2$ and CO are more abundant at the edge of the CSE than predicted by previous models due to molecular shielding from photodissociation.

\item The photodissociation of \ce{N2} and CO affects the chemistry of most species in the outer CSE.
Following its improved treatment here, the transition zones of N$_2$ $\to$ N and CO $\to$ \ce{C+} $\to$ C shift towards the outer edge of the \textcolor{black}{envelope}.
The abundances of some species formed from N are reduced in the outer CSE.
This induces large changes in the predicted column densities (factor of 10) of some species (e.g., \ce{C_nN} and \ce{C_nN-} carbon chains) and the radii of their peak abundances.



\item \textcolor{black}{Predictions are made for column densities
and peak radii of molecules whose abundances are sensitive to \ce{N2}
and CO photodissociation and which could conceivably be detected.
These may be verified by observations using high resolution and high sensitivity telescopes, especially ALMA and PdBI.}

\item \textcolor{black}{The abundances obtained from models which
treat the full angularly-resolved spherically-symmetric radiation
field incident on the outer CSE are very similar to those found in
plane-parallel models in which the radiation is incident normally but
with the intensity reduced by half. The proposed iterative method for
implementing molecular shielding functions is most efficient when
assuming that the molecule is fully shielded at the beginning of the
calculation.  }

 \end{enumerate}


Other astrochemical models simulating the CSEs of AGB stars or other environments may show a similar sensitivity \textcolor{black}{using a} more realistic treatment of self-shielding and spherically-symmetric radiation as \textcolor{black}{was} found in this model of IRC +10216.

The photodissociation processes \textcolor{black}{for} \ce{N2} and CO (both the unshielded rate as well as the depth dependence) are now very well understood,  so any major discrepancies between observations and models  \textcolor{black}{are likely} due to other assumptions in the model.



\begin{acknowledgements}   
\textcolor{black}{The authors would like to thank the anonymous referee for his/her valuable comments and constructive suggestions.} X. Li is delighted to thank TJM for his hospitality during a two-week visit to Queen's University Belfast (QUB) in 2013, and thank Dr. Markus Schmalzl and Prof. Xander Tielens for some stimulating discussions. 

Astrochemistry in Leiden is supported by the Netherlands Research School for Astronomy (NOVA), by a Spinoza grant and grant 648.000.002 from the Netherlands Organisation for Scientific Research (NWO), and by the European Community's Seventh Framework Programme FP7/2007-2013 under grant agreements 291141 (CHEMPLAN) and 238258 (LASSIE). Astrophysics at QUB is supported by a grant from the STFC. \textcolor{black}{C. W. acknowledges support from the NWO (program number 639.041.335).}

\end{acknowledgements}
\bibliographystyle{aa}
\bibliography{cse}

\begin{appendix}
\section{Spherically-symmetric model} \label{sec:appendix_3D_model}
\placefigurecDmodel        \label{placefigure3Dmodel} 
\placefigureNbrceta        \label{placefigureNbrceta} 
In this appendix we describe the SS model employed in this work for the computation of the angle-dependent column densities of species at each radius. The SS model is shown in Fig. \ref{fig3D_model}. The star is located at point S and ejects dust and molecules to various shells with radii $r_i$ (not equidistant). Since the star itself is too cool to generate photons that can induce photodissociation of molecules (in our case, IRC +10216, $\sim 2330$ K), one only needs to consider photons from the interstellar radiation field. For a specific position P at radius $r_i$, suppose $\vartheta$ \textcolor{black}{ is the angle between the direction of the incoming ray and the radius vector from the central star.} Generally, the column density, $N$, of a species is defined as
\begin{align}
N \equiv \int_0^{\infty} \! n \, \mathrm{d}z ~~ \text{cm}^{-2}, 
\end{align}
where $n$ is the number density in units of cm$^{-3}$ and $z$ is the integral path length in units of cm. In the present study, the total column density, $N(r_i,\vartheta)$, of a species integrated along a ray from point P to infinity is given by two cases:
\begin{itemize}  
	\item[(1)] $0$ $\leq$ $\vartheta$ $\leq$ $\pi/2$, for example, $\overline{\text{PA}}$, yielding
	\begin{align} \label{eq:N_r_theta}
&   N_{\overline{\text{PA}}}(r{_i},\vartheta)=\dfrac{1}{2}\left(\sqrt{r_{i+1}^{2}-r_{i}^{2}\sin^{2}\vartheta}-r_{i}\cos\vartheta\right) \left[ n(i)+ n(i+1)\right]                    \notag          \\ 
& ~~~~~~~~~~~~~~~~~~~~ + \textcolor{black}{   \dfrac{1}{2}\sum\limits_{j=i+1}^{n-1}\left(\sqrt{r_{j+1}^{2}-r_{i}^{2}\sin^{2}\vartheta}-\sqrt{r_j^2 - r_{i}^{2}\sin^{2}\vartheta}\right) }  \notag  \\
& ~~~~~~~~~~~~~~~~~~~~~\textcolor{black}{  [n(j)+n(j+1)]   } ~~~~\text{cm}^{-2},     
    \end{align}  
where $n(i)$ is the number density of the molecule at radius $r_i$.
	\item[(2)] $\pi/2$ $<$ $\vartheta$ $\leqslant$ $\pi$, for instance $\overline{\text{BP}}$. Since we assume the model is spherically symmetric, one may simplify the calculations by finding a point C that satisfies $\overline{\text{SC}}$ $\bot$ $\overline{\text{BD}}$, then get the column density along $\overline{\text{BP}}$ with  
	\begin{align} \label{eq:N_r_theta2}	
&   N_{ \overline{\text{PB}}} \left(r{_i}, \vartheta \right) = 2 N_{\overline{\text{CB}}} \left( r{_i},\pi /2 \right) - N_{\overline{\text{PD}}} \left(r{_i},\pi-\vartheta \right)~~\text{cm}^{-2}.            
    \end{align} 	 
Both $N_{\overline{\text{CB}}} \left(r{_i},\pi /2 \right)$ and $N_{\overline{\text{PD}}} \left(r{_i},\pi-\vartheta \right)$ can be conveniently calculated by case (1). Then, it is necessary to interpolate an abundance between the radially-gridded values in our model for points near the star, such as C.
\end{itemize}

Figure \ref{figNbrceta} shows $N(r_i,\vartheta)$ for molecular N$_2$. According to our calculations, the major contribution of photons that induce photodissociation of N$_2$ come from the area within the dashed line ($N(r_i,\vartheta) < 1.0 \times 10^{-16}$ cm$^{-2}$), where $r_i > 1.6 \times 10^{17}$ cm and $\vartheta < 3\pi/4$. \textcolor{black}{ Deeper in the envelope,} the fractional abundance of N$_2$ stays constant because it  \textcolor{black}{is} fully shielded.


\section{Implementing the molecular shielding functions} \label{sec:appendix_loop}
\placefigureLoop     \label{placefigureLoop}    
In this Appendix we introduce the new procedure employed in the current work for the implementation of molecular shielding functions. Here we take N$_2$ as an example for the demonstration. The method has five steps.
\begin{enumerate}[Step 1] 
	\item Set up the initial condition. Suppose N$_2$ is fully shielded everywhere in the CSE, then we can set $k^\text{initial}(r_i)\simeq 0$ for each radius, $r_i$, which means the corresponding shielding functions from dust and molecules are zero, i.e., $\Theta_\text{dust}^\text{initial}(r_i,\vartheta)=0$, and $\Theta_\text{mol}^\text{initial}(r_i,\vartheta)=0$. In all cases, we consider 201 incident rays as a function of $\vartheta$ between $0$ and $\pi$, where $\vartheta=0$ gives the normal direction.  
	\item  Loop 1: compute the CSE chemistry for the first time, employing $k^\text{initial}(r_i)$ prepared from Step 1. The calculated N$_2$ abundance generates column densities, $N(r_i,\vartheta)$, and corresponding dust and molecular shielding functions,  $\Theta_\text{dust}^\text{loop1}(r_i,\vartheta)$, and $\Theta_\text{mol}^\text{loop1}(r_i,\vartheta)$, and new photodissociation rates $k^\text{loop1}(r_i)$. 
	\item  Loop 2: repeat the simulations of the CSE chemistry, but employing $k^\text{loop1}(r_i)$ produced in Step 2 for each radius. Using the same method, generate another set of photodissociation rates, $k^\text{loop2}(r_i)$. 
	\item  Compare the fractional abundances of N$_2$ at each radius for Loop 1 and  Loop 2, to see if they are the same. If no changes are found, go to Step 5, otherwise go to Step 3, continuing the calculations.
	\item  Stop the calculation. The outputs at this step are the final results.
\end{enumerate}
The simulations can also approach convergence by assuming N$_2$ is unshielded in Step 1, namely $k^\text{initial}(r_i)=k^\text{0}$ at each radius. In this case, the fractional abundance of N$_2$ will steeply decrease after starting the evolution of the chemical network, then approach the converged results quickly, as illustrated in Fig.~\ref{figloop}. The same converged results can be obtained \textcolor{black}{from} the two \textcolor{black}{possible} assumptions \textcolor{black}{in} Step 1. In practice, the fractional abundances of both N$_2$ and CO (as well as all other species) reach their converged abundances after $\sim$ 15 loops. In practice, starting the calculation assuming full shielding in Loop 1 is most efficient. 

The iterative method proposed here may be used in other models, where output variables are also required as inputs.
    
\end{appendix}
\end{document}